\documentclass[twocolumn]{aastex631}
\usepackage{color}
\usepackage[titletoc]{appendix}
\usepackage{amsmath}
\usepackage{amssymb}
\usepackage{mathtools}
\usepackage{upgreek}
\usepackage{float}
\usepackage{comment}
\usepackage{enumitem}
\usepackage{natbib}
\usepackage{graphicx}
\usepackage{bm}
\usepackage{totcount}
\usepackage{multirow}

\newtotcounter{citnum} %From the package documentation
\def\oldbibitem{} \let\oldbibitem=\bibitem
\def\bibitem{\stepcounter{citnum}\oldbibitem}

\shortauthors{Millholland et al.}
\shorttitle{Edge-of-the-multis}

\begin{document} 

\title{\small{Edge-of-the-Multis: Evidence for a Transition in the Outer Architectures of Compact Multi-Planet Systems}}

\author[0000-0003-3130-2282]{Sarah C. Millholland}
\affiliation{Department of Physics, Massachusetts Institute of Technology, Cambridge, MA 02139, USA}
\affiliation{MIT Kavli Institute for Astrophysics and Space Research, Massachusetts Institute of Technology, Cambridge, MA 02139, USA}
\affiliation{Department of Astrophysical Sciences, Princeton University, Princeton, NJ 08544, USA}
\email{sarah.millholland@mit.edu}

\author[0000-0002-5223-7945]{Matthias Y. He}
\affiliation{Department of Physics, University of Notre Dame, Notre Dame, IN 46556, USA}
\affiliation{Department of Astronomy \& Astrophysics, The Pennsylvania State University, University Park, PA 16802, USA}

\author[0000-0003-1848-2063]{Jon K. Zink}
\altaffiliation{NASA Sagan Fellow}
\affiliation{Department of Astronomy, California Institute of Technology, Pasadena, CA 91125, USA}

\begin{abstract}
Although the architectures of compact multiple-planet systems are well-characterized, there has been little examination of their ``outer edges'', or the locations of their outermost planets. Here we present evidence that the observed high-multiplicity Kepler systems truncate at smaller orbital periods than can be explained by geometric and detection biases alone. To show this, we considered the existence of hypothetical planets orbiting beyond the observed transiting planets with properties dictated by the ``peas-in-a-pod'' patterns of intra-system radius and period ratio uniformity.  We evaluated the detectability of these hypothetical planets using (1) a novel approach for estimating the mutual inclination dispersion of multi-transiting systems based on transit chord length ratios and (2) a model of transit probability and detection efficiency that accounts for the impacts of planet multiplicity on completeness. 
% Add ``conservative''
Under the assumption that the ``peas-in-a-pod'' patterns continue to larger orbital separations than observed, we find that $\gtrsim35\%$ of Kepler compact multis should possess additional detected planets beyond the known planets, constituting a $\sim7\sigma$ discrepancy with the lack of such detections. These results indicate that the outer ($\sim100-300$ days) regions of compact multis experience a truncation (i.e. an ``edge-of-the-multis'') or a significant breakdown of the ``peas-in-a-pod'' patterns, in the form of systematically smaller radii or larger period ratios. We outline future observations that can distinguish these possibilities, and we discuss implications for planet formation theories.
\\
\end{abstract}

\section{Introduction}
\label{sec: Introduction}

Short-period, sub-Neptune-sized planets are one of the most prevalent types of exoplanets in the local Galaxy. Discovered in abundance by NASA's Kepler mission \citep{2010Sci...327..977B}, these planets are frequently found in multiple-planet systems with orbital periods ranging from days to months \citep{2011ApJS..197....8L, 2014ApJ...784...44L, 2014ApJ...784...45R, 2014ApJ...790..146F}. Over the last decade, the architectures of these compact-multiple planet systems (``compact multis'') have been characterized in detail. The observed planets have low eccentricities \citep[e.g.][]{2015ApJ...808..126V,2016PNAS..11311431X}, low inclinations \citep[e.g.][]{2012ApJ...761...92F, 2014ApJ...790..146F}, and tight orbital spacings that are generally not near resonances \citep{2011ApJS..197....8L, 2014ApJ...790..146F, 2015ARA&A..53..409W}. Within a given system, the period ratios, planet radii, and planet masses tend to be significantly more uniform than would be expected by random chance -- a set of patterns known collectively as the ``peas-in-a-pod patterns'' or ``intra-system uniformity''  (\citealt{2018AJ....155...48W}, \citealt{2017ApJ...849L..33M}; for a review, see \citealt{2022arXiv220310076W}). 

The statistical properties of the \textit{observed} compact multis have in turn revealed corresponding properties of the \textit{underlying} distribution of planetary systems (i.e. including undetected planets) through the help of forward models. For instance, the fraction of stars with multiple planets interior to $\sim 1$ AU is estimated at $\sim30\%-70\%$ \citep{2018ApJ...860..101Z, 2018AJ....156...24M, 2019MNRAS.483.4479Z, 2019MNRAS.490.4575H}, with the occurrence increasing for cooler stars \citep{2012ApJS..201...15H, 2015ApJ...814..130M, 2020AJ....159..164Y, 2021AJ....161...16H}. There is evidence that the eccentricities and inclinations correlate with the intrinsic planet multiplicity \citep{2018ApJ...860..101Z, 2020AJ....160..276H, 2021ARA&A..59..291Z} and that the inclinations are both relatively small overall and not bifurcated into dynamically cool and dynamically hot sub-populations \citep{2021AJ....162..166M}. Finally, forward models have shown that the patterns of intra-system uniformity in planetary sizes and orbital spacings must be present to a strong degree in the underlying population \citep{2019MNRAS.490.4575H, 2020AJ....160..276H, 2020AJ....159..281G, 2021A&A...656A..74M}, indicating that these patterns cannot be explained by detection biases and require astrophysical origins.

There is an additional feature of compact multis that, relative to the properties mentioned above, has received comparatively little attention. That is their inner and outer edges, which can be summarized statistically as the distributions of orbital periods of the innermost and outermost planets in the systems. The edges of compact multis are not well understood from either an observational or theoretical perspective. However, these features (specifically those of the \textit{underlying} population of systems) are crucial for a complete characterization of short-period tightly-packed systems.

The inner edges are the simpler of the two from an observational perspective, since the geometric and detection biases are comparatively minimal. A simple calculation based on a sample of Kepler systems with four or more observed planets (to be defined later in this paper) shows that the distribution of observed innermost planet periods has a median of 3.9 days with the 16th-84th percentile interval equal to [2.2, 7.0] days. After accounting for biases, \cite{2018AJ....156...24M} showed that the underlying distribution of inner edges likely peaks at slightly longer orbital periods of about $\sim10$ days. This is consistent with being a signature from the protoplanetary disk inner edges \citep[e.g.][]{2007prpl.conf..539M}, which may limit where planets are able to form \textit{in situ} \citep[e.g.][]{2017ApJ...842...40L} or may act as a planet trap that halts inward migration \citep[e.g][]{2007ApJ...654.1110T, 2017MNRAS.470.1750I}.

The outer edges are considerably more difficult to understand because they are heavily impacted by geometric and detection biases. Specifically, the transit probability and transit signal-to-noise ratio decrease with increasing orbital period as $p_{\mathrm{trans}} \sim P^{-2/3}$ and ${\mathrm{SNR} \sim P^{-1/3}}$, respectively. Moreover, transit surveys require a minimum number of transits for detection (three for the Kepler pipeline), which, together with the observation baseline and duty cycle, limits the largest detectable orbital period. This upper limit is $P \lesssim 500$ days for the Kepler prime mission. For a sample of Kepler systems with four or more observed planets (the same sample as discussed in the previous paragraph), the distribution of outermost planet periods has a median of 40.6 days with the 16th-84th percentile interval equal to [16.3, 76.9] days. Naively speaking, this distribution already seems inconsistent with expectations, since almost all of the outermost periods are more than an order of magnitude smaller than the upper limit of $\sim$500 days. However, such comparisons are not very meaningful until the relevant geometric and detection biases have been thoroughly accounted for.

In this work, we examine the outer edges of compact multis discovered by the Kepler prime mission.\footnote{Although K2 and TESS have also discovered some compact multis, we restrict our focus to the large sample of systems discovered by the Kepler prime mission, since its homogeneous coverage out to large orbital periods is required for this study.} Our main goal is to understand whether the outer edges of the \textit{observed} systems are consistent with sculpting purely from geometric and detection biases acting upon an \textit{underlying} population of systems that extend out to (and potentially beyond) the largest periods probed by the Kepler photometry. Alternatively, if we find that the observed systems truncate at smaller orbital periods than required by their detection, then this could indicate either that (1) the underlying planets in compact multis are present only within some restricted range of orbital separations and/or (2) there are other significant changes in the planet properties (e.g. periods, radii) at larger orbital separations. To explore this, we will consider the existence of hypothetical planets orbiting beyond the outermost planets in Kepler multis, and we will estimate the number of these hypothetical planets that we would expect to be detectable. 

It is important to note that this investigation of the outer edges of compact multis has consequences beyond demographics; specifically, the outer edges are signatures of planet formation and dynamical evolution. Just as the inner edges of compact multis are likely relics from the disk inner edges, a possible truncation in the outer systems could be the result of disk migration traps \citep[e.g.][]{2022arXiv220205342Z}, planet formation in pebble rings \citep[e.g][]{2014ApJ...780...53C}, or dynamical perturbations from exterior giant planets \citep[e.g.][]{2018MNRAS.478..197P}, among other possibilities. We will review such theoretical considerations in Section \ref{sec: discussion}. However, it is helpful to keep this motivation in mind from the outset. 

This paper is organized as follows. We begin with a heuristic demonstration of the detectability of potential transiting outer planets, as a motivation for the more detailed calculations in subsequent sections (Section \ref{sec: heuristic calculation}). We then investigate the geometric aspects of the problem and develop a new method for estimating a system's mean inclination and mutual inclination dispersion (Section \ref{sec: transit chord ratio method}). Based on these estimates, we define a model for the transit probability, as well as a detection efficiency model (Section \ref{sec: transit probabilities and detection probabilities}). We then validate these approaches on populations of simulated planets (Section \ref{sec: Validation Tests with SysSim}). Moving onto observed systems, we estimate the number of hypothetical additional outer planets that we would expect to be transiting and detectable in Kepler multiple-planet systems (Section \ref{sec: hypothetical planet experiments}). Finally, we discuss implications of our results for planet formation theories (Section \ref{sec: discussion}).

\section{Heuristic Calculation}
\label{sec: heuristic calculation}

\begin{figure}
\centering
\epsscale{1.2}
\plotone{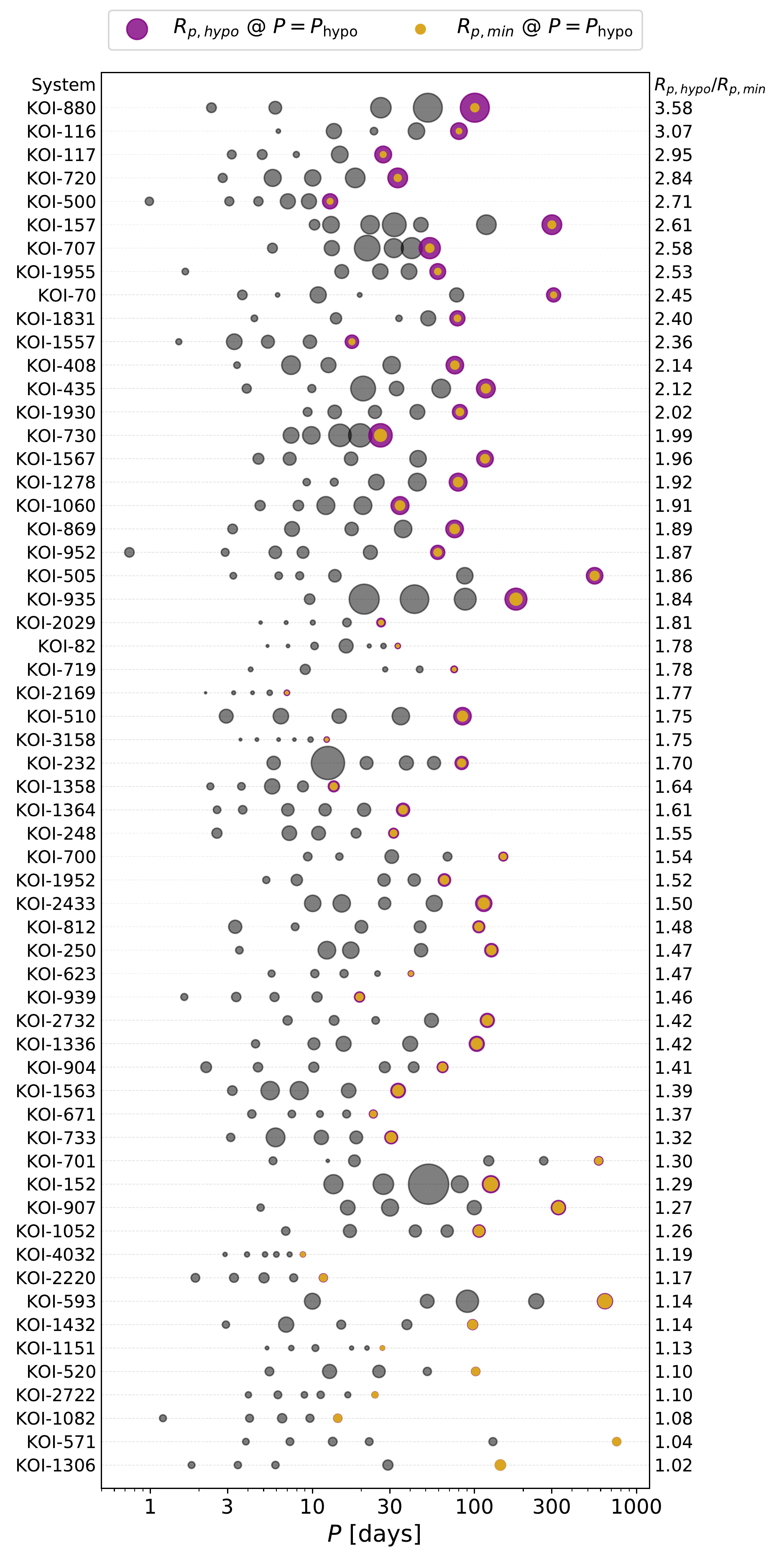}
\caption{Architectures of Kepler systems with four or more transiting planets, where all planets in the system are smaller than $8 \ R_{\oplus}$. The dot size is proportional to the planet radius. Observed planets are indicated in gray, while the ``hypothetical planets'' with $R_{p,\mathrm{hypo}} = R_{p,\mathrm{out}}$ and $P_{\mathrm{hypo}}/P_{\mathrm{out}} = P_{\mathrm{out}}/P_{\mathrm{2nd-out}}$ are shown in purple. Yellow dots indicate the minimum planet radius, $R_{p,\mathrm{min}}$, necessary for $\mathrm{SNR}>7$ at $P=P_{\mathrm{hypo}}$. The systems are reverse-ordered by the ratio, $R_{p,\mathrm{hypo}}/R_{p,\mathrm{min}}$, as shown at right. In every system, $R_{p,\mathrm{hypo}}$ is greater than $R_{p,\mathrm{min}}$. } 
\label{fig: Kepler_multis_system_lineup}
\end{figure}

We begin with a heuristic exploration of the detectability of potential transiting outer planets in Kepler multi-planet systems. This offers a simplified preview of later detailed calculations. We consider a sample of 64 Kepler systems with four or more transiting planets. (The details of the sample selection will be provided in Section \ref{sec: sample selection}.) We imagine that each system has an additional transiting outer planet with properties consistent with the ``peas-in-a-pod'' patterns. Thus, we set the radii and periods of these ``hypothetical planets'' to $R_{p,\mathrm{hypo}} = R_{p,\mathrm{out}}$ and $P_{\mathrm{hypo}}/P_{\mathrm{out}} = P_{\mathrm{out}}/P_{\mathrm{2nd-out}}$, where $P_{\mathrm{out}}$ and $R_{p,\mathrm{out}}$ are the period and radius of the outermost observed planet, and $P_{\mathrm{2nd-out}}$ is the period of the second outermost observed planet.

Next, we compare $R_{p,\mathrm{hypo}}$ to the minimum planet radius, $R_{p,\mathrm{min}}$, that the hypothetical planets would require in order to be detected with signal-to-noise ratio, ${\mathrm{SNR}>\mathrm{SNR_{min}}=7}$, which is approximately equal to the detection threshold of the Kepler pipeline. We define the SNR as \citep{2012PASP..124.1279C}
\begin{equation}
\mathrm{SNR}=\frac{(R_p/R_{\star})^2}{\mathrm{CDPP}_{\mathrm{eff}}}\sqrt{\frac{t_{\mathrm{obs}}f_0}{P}}.
\end{equation}
Here, $t_{\mathrm{obs}}$ is the time interval over which data were collected ($\sim$4 years in most cases), $f_0$ is the duty cycle, and $\mathrm{CDPP}_{\mathrm{eff}}$ is the effective combined differential photometric precision,
\begin{equation}
\mathrm{CDPP}_{\mathrm{eff}}=\mathrm{CDPP}_{6\,\mathrm{hr}}\sqrt{\frac{6\,\mathrm{hr}}{T}},
\end{equation}
where $T \propto P^{-1/3}$ is the transit duration (see equation \ref{eq: Tdur} for the full expression).
In Section \ref{sec: detection probabilities}, we will use a more advanced definition of the SNR, but this definition suffices for the purposes of this illustration. Given that $\mathrm{SNR} \propto P^{-1/3}{R_p}^2$, we can calculate the minimum planet radius, $R_{p,\mathrm{min}}$, for which ${\mathrm{SNR} > \mathrm{SNR_{min}} = 7}$ at $P = P_{\mathrm{hypo}}$ using a scaling to $\mathrm{SNR_{out}}$, the SNR of the outermost observed planet in the system,
\begin{equation}
R_{p,\mathrm{min}} = R_{p,\mathrm{out}}\left(\frac{\mathrm{SNR_{min}}}{\mathrm{SNR_{out}}}\right)^{1/2}\left(\frac{P_{\mathrm{hypo}}}{P_{\mathrm{out}}}\right)^{1/6}.
\end{equation}

Figure \ref{fig: Kepler_multis_system_lineup} shows the architectures of Kepler systems with four or more planets, along with dots indicating the radii $R_{p,\mathrm{hypo}}$ and $R_{p,\mathrm{min}}$ at period $P_{\mathrm{hypo}}$. For the purposes of the visualization, we include only systems for which all planets in the system have $R_p < 8 \ R_{\oplus}$, leaving out 5 systems. Remarkably, in all 59 systems, $R_{p,\mathrm{hypo}} > R_{p,\mathrm{min}}$. In other words, the hypothetical planet's nominal radius is larger than the minimum radius required for $\mathrm{SNR}>7$ in every single case. This result can be seen by the purple outlines around the yellow dots in Figure \ref{fig: Kepler_multis_system_lineup} and the $R_{p,\mathrm{hypo}}/R_{p,\mathrm{min}}$ ratios on the right-hand-side. If we increase $P_{\mathrm{hypo}}$, then $R_{p,\mathrm{hypo}} > R_{p,\mathrm{min}}$ remains to be generally true but not in every case. For instance, if we set $P_{\mathrm{hypo}} = 300$ days in all systems, then $R_{p,\mathrm{hypo}} > R_{p,\mathrm{min}}$ in 45 out of 59 systems.

To summarize, this brief experiment has demonstrated the following: If additional transiting planets exist beyond the known transiting planets, and if they have properties in line with expectations from the ``peas-in-a-pod'' patterns, then we would generally expect them to be detectable. In the remainder of the paper, we explore this simple premise in greater detail. We consider similar experiments involving hypothetical outer planets, but we consider both transit probabilities and detection probabilities and formulate detailed expectations of the frequency of such planets we would expect to be transiting and detectable. In the next section, we begin with the key geometric aspect of the problem: mutual orbital inclinations.

\section{Mutual Inclination Estimation using Transit Chord Ratios}
\label{sec: transit chord ratio method}

\subsection{Overview of the transit chord ratio method}

One of our subsequent objectives is to quantify the likelihood that planets orbiting beyond the known planets in a multi-transiting system are also transiting. In order to do this, we must first obtain estimates of the mean inclination and inclination dispersion. Here we develop a new procedure called the ``transit chord ratio method'', which uses the ratios of the transit chord lengths of pairs of planets in the same system \citep{2010ApJ...725.1226S} as a means of constraining the planets' inclinations. This has been done at the population level \citep[e.g.][]{2012ApJ...761...92F, 2014ApJ...790..146F} but not the level of individual systems. One benefit of working with ratios of transit chord lengths (rather than each planet's individual transit duration) is that it significantly reduces the impact of the degeneracy between the stellar density and impact parameter, which are difficult to accurately measure with Kepler long cadence photometry \citep{2020AJ....160...89P}.

It is important to note that transit chord ratios are sensitive specifically to sky-plane inclinations, the inclinations between the planetary orbit planes and the sky plane. Accordingly, they cannot directly constrain the dispersion of the inclinations between the planetary orbits and the invariable plane (the plane perpendicular to the system's total angular momentum vector). Moreover, since the transit chord lengths are symmetric on the two stellar hemispheres bisected by the $b=0$ chord, the method cannot determine whether two planets are transiting on the same hemisphere or opposite ones. Despite these limitations, the method is still capable of providing useful estimates of the mean inclination and inclination dispersion of multi-transiting systems, as we will soon demonstrate.

In the limit $R_p \ll R_\star \ll a$, a planet's transit duration is given by
\begin{equation}
T = \frac{2 R_{\star} \sqrt{1-b^2}}{v_{\mathrm{mid}}}, 
\label{eq: Tdur}
\end{equation}
where $R_{\star}$ is the stellar radius, $b$ is the dimensionless impact parameter, and $v_{\mathrm{mid}}$ is the sky-projected orbital velocity at mid-transit. Here we will make the simplifying assumption of circular orbits,\footnote{This assumption will slightly increase our estimates of the inclination dispersion because any eccentricity-driven spread of the transit durations from the circular orbit expectations will be interpreted as a spread in impact parameters. In practice, this is a small effect due to the nearly circular orbits of planets in compact multis.} such that $v_{\mathrm{mid}} = na$, where $n = 2\pi/P$ is the mean-motion and $a$ is the semi-major axis. The ratio of transit chord lengths ($T v_{\mathrm{mid}}$) of a pair of planets ($j$ and $k$ with $j < k$) in the same system is 
\begin{equation}
\begin{split}
\xi_{j,k} &\equiv \left(\frac{T_k}{T_j}\right)\left(\frac{P_{k}}{P_{j}}\right)^{-1/3} = \sqrt{\frac{1-(a_k/R_\star)^2\cos{^2{i_k}}}{1-(a_j/R_\star)^2\cos{^2{i_j}}}}\\
&= \sqrt{\frac{1-(\frac{1}{3}G\pi^{-1}\rho_\star{P_k}^2)^{2/3}\cos{^2{i_k}}}{1-(\frac{1}{3}G\pi^{-1}\rho_\star{P_j}^2)^{2/3}\cos{^2{i_j}}}},
\label{eq: transit chord ratio}
\end{split}
\end{equation}
where on the right-hand side of the first line, we have made the substitution $b R_{\star} = a \cos{i}$, with $i$ the inclination between the planet's orbit plane and the sky plane. In the final line, we expressed $a/R_\star$ in terms of the orbital period and stellar density, $\rho_\star$. Since $T$ and $P$ are well-measured from transit data, the observed ratios $(\xi_{j,k})_{\mathrm{obs}} = (T_k/T_j)(P_{k}/P_{j})^{-1/3}$ can provide constraints on the inclinations \citep{2014ApJ...790..146F}. The uncertainties on $(\xi_{j,k})_{\mathrm{obs}}$ can be calculated through standard error propagation using the reported uncertainties on $T$ and $P$. 

\begin{figure}
\centering
\epsscale{1.2}
\plotone{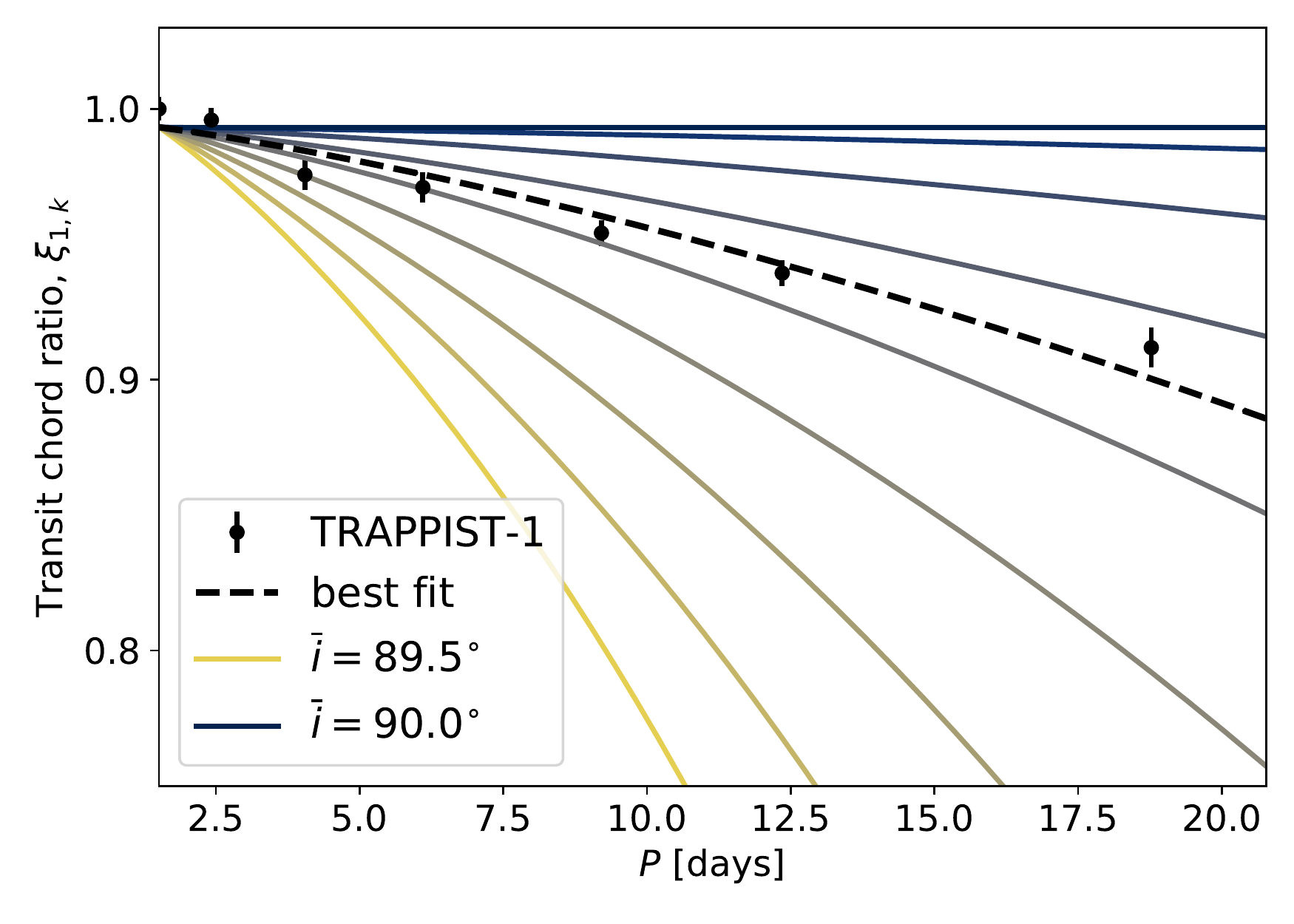}
\caption{Transit chord ratios for planets in the TRAPPIST-1 system. The black dots with error bars indicate the observed ratios,  $(\xi_{1,k})_{\mathrm{obs}} = (T_k/T_1)(P_{k}/P_{1})^{-1/3}$. The solid curves represent coplanar models, $(\xi_{1,k})_{\mathrm{mod}}$, with different inclinations, ranging from $\overline{i} = 89.5^\circ$ to $\overline{i} = 90^\circ$ in steps of $0.0625^\circ$. The black dashed line indicates the model curve with best-fit parameters $\overline{i} = 89.781^\circ$ and $\gamma = 0.993$.} 
\label{fig: TRAPPIST-1_example}
\end{figure}

Consider the ratio of transit chord lengths of each planet in a multi-planet system with respect to that of the innermost planet, $\xi_{1,k}$, where $k = 1, 2, ..., N$. For a perfectly coplanar system with constant sky-plane inclination $\overline{i}$, data points of ($P$, $\xi_{1,k}$) would trace out a smooth and monotonic curve. Examples of these curves are depicted in Figure \ref{fig: TRAPPIST-1_example} and are calculated by setting the inclinations to $\overline{i}$ in equation \ref{eq: transit chord ratio}. For a set of observed $(\xi_{1,k})_{\mathrm{obs}}$ values in a system, we can identify a model curve, $(\xi_{1,k})_{\mathrm{mod}}$, that best fits the observations. The spread of the $(\xi_{1,k})_{\mathrm{obs}}$ measurements around the best-fit curve indicates the degree of sky-plane inclination dispersion about the mean sky-plane inclination.

When fitting the model curve, we allow $(\xi_{1,k})_{\mathrm{mod}}$ to be modified by a vertical scaling factor, $\gamma$, which accounts for the fact that $i_1$ may not be near $\overline{i}$. In other words, $(\xi_{1,k})_{\mathrm{mod}}$ need not pass through unity at $P = P_1$, since this can otherwise bias the fit. We use non-linear least squares to find the best-fit values of $\overline{i}$ and $\gamma$ that minimize the sum of the squared residuals of $(\xi_{1,k})_{\mathrm{mod}} - (\xi_{1,k})_{\mathrm{obs}}$. We require $\overline{i} \leq 90^{\circ}$ in the fit, which is not a loss of generality due to the symmetry of the transit chord lengths on the two stellar hemispheres bisected by the $b=0$ chord. The best-fit values of $\overline{i}$ and $\gamma$ uniquely yield $i_1$, which we then use to solve for the other planets' inclinations from equation \ref{eq: transit chord ratio}. Finally, we measure the sky-plane inclination dispersion as the root mean square of $\Delta i_k = i_k - \overline{i}$,
\begin{equation}
\sigma_i = \sqrt{\frac{1}{N}\sum_{k=1}^{N} (i_k - \overline{i})^2}.
\label{eq: sigma_i}
\end{equation}

We illustrate the transit chord ratio method using the TRAPPIST-1 system \citep{2017Natur.542..456G} as an example. With well-determined parameters for all seven planets \citep{2021PSJ.....2....1A}, the TRAPPIST-1 system allows us to compare the constraints on the mean inclination $\overline{i}$ and dispersion $\sigma_i$ from the transit chord ratio method to those obtained from independent techniques. We adopt the values of $\rho_{\star}$ and each planet's $T$ and $P$ from \cite{2021PSJ.....2....1A}. Figure \ref{fig: TRAPPIST-1_example} shows the observed ratios, $(\xi_{1,k})_{\mathrm{obs}} = (T_k/T_1)(P_{k}/P_{1})^{-1/3}$, along with the model curve with best-fit parameters, $\overline{i} = 89.781^\circ$ and $\gamma = 0.993$. The observed values are tightly distributed around the best-fit curve, yielding a dispersion equal to $\sigma_i = 0.105^\circ$. The calculated estimates of $\overline{i}$ and $\sigma_i$ agree well with those derived from the photodynamical model in \cite{2021PSJ.....2....1A}, where the mean and dispersion of the $i$ values are $89.783^\circ\pm0.032^\circ$ and $0.053^\circ\pm0.034^\circ$. 

The TRAPPIST-1 system is exceptionally close to coplanar. By experimenting with synthetic systems with larger inclination dispersion (such as some of the systems that will be described in the next section), we identified that when $\sigma_i \gtrsim 0.4^{\circ}$, the fit sometimes forces $\overline{i}$ to nearly its maximum possible value, $90^{\circ}$. This occurs when several planets have $|b_k| < |b_1|$, which leads to the corresponding ratios $(\xi_{1,k})_{\mathrm{obs}} > 1$. The model attempts to capture these points in the fit (along with those with $(\xi_{1,k})_{\mathrm{obs}} < 1$) by forcing $\overline{i}$ to $90^{\circ}$, which corresponds to a horizontal line of $(\xi_{1,k})_{\mathrm{mod}}$ vs. $P$. To remedy these cases, we utilize a second estimate of $\overline{i}$ that is derived by first estimating the inclination for each planet directly from its measured transit duration (equation \ref{eq: Tdur}) and then taking the mean over all planets, $\overline{i}_{\mathrm{circ}}$. For cases where the transit chord ratio method yields $\sigma_i > 0.4^{\circ}$ and $\overline{i} > 89.99^{\circ}$, we replace the original value of $\overline{i}$ with $(\overline{i} +  \overline{i}_{\mathrm{circ}})/2$. Whenever we replace $\overline{i}$, we also recalculate $\sigma_i$ using equation \ref{eq: sigma_i}.  

\subsection{Validation using simulated planetary systems}
\label{sec: transit chord ratio method validation using SysSim}

We can examine the accuracy of the transit chord ratio method using simulated planetary systems, for which we can compare the derived mean inclination and inclination dispersion to the system's true values. We consider simulated planetary systems from the SysSim (short for ``Planetary Systems Simulator'') forward modeling framework\footnote{The core SysSim code and the specific forward model explored in this study are available at \href{https://github.com/ExoJulia/ExoplanetsSysSim.jl}{https://github.com/ExoJulia/ExoplanetsSysSim.jl} and \href{https://github.com/ExoJulia/SysSimExClusters}{https://github.com/ExoJulia/SysSimExClusters}.} \citep{2018AJ....155..205H, 2019AJ....158..109H, 2019MNRAS.490.4575H, 2020AJ....160..276H, 2021AJ....161...16H}. SysSim is a tool used to generate simulated planetary systems according to flexible statistical models, the parameters of which are determined using a calibration to summary statistics of the observed Kepler planet population (including the observed distributions of multiplicities, orbital periods and period ratios, transit depths and depth ratios, etc.). 

It is important to distinguish between two different planetary populations that are generated within a SysSim forward model. First, there is the ``physical catalog'', the underlying population of planetary systems, which is drawn directly from the statistical model. Second, there is the ``observed catalog'', which is obtained by passing the physical catalog through a simulated Kepler detection pipeline and determining which planets would be detected and labeled as planet candidates during the automated vetting process. Once the best-fit parameters of the forward model have been identified (see e.g. \citealt{2019AJ....158..109H, 2019MNRAS.490.4575H} for details), the observed catalog closely resembles the population of observed Kepler systems.

In this study, we will consider simulated planetary systems from the latest edition of SysSim: the ``maximum AMD model'' \citep{2020AJ....160..276H}. This model is based on the argument that a system's long-term orbital stability is closely related to its angular momentum deficit (AMD), the difference between the total orbital angular momentum of the system and what it would be if all orbits were circular and coplanar \citep[e.g.][]{2017A&A...605A..72L}. The key assumption of the model is that all systems have the critical AMD for stability. Although we do not expect this to be true for all systems in reality, the model offers an intuitive framework for assigning system orbital properties and has been shown to reproduce many aspects of the Kepler data \citep{2020AJ....160..276H}, including Kepler planet Transit Duration Variations (TDVs), which probe mutual orbital inclinations \citep{2021AJ....162..166M}. In this work, we will use pairs of physical and observed catalogs generated from the maximum AMD model. Each catalog pair corresponds to a different set of model parameters sampled from the posterior distributions derived by \cite{2020AJ....160..276H}.

We utilize ten physical/observed catalog pairs and consider only the systems with four or more detected planets (``4+ systems''), equaling 374 systems total. The observed catalogs contain estimates of $T$ and $P$ of the simulated observed planets with realistic measurement precision. They also contain the ``true'' values of the stellar density (i.e., assuming perfect measurement precision). In order to simulate realistic uncertainties of the stellar density measurements, we utilize the fractional uncertainties, $f_{\rho_\star} = \sigma_{\rho_\star}/\rho_\star$, of Kepler host stars from the Gaia-Kepler Stellar Properties Catalog \citep{2020AJ....160..108B}, where we consider only the subset of host stars with four or more observed planets. The median $f_{\rho_\star}$ of this sub-sample is 9\%. For each $\rho_{\star,\mathrm{true}}$ value in the SysSim catalog, we draw a random fractional uncertainty, $f_{\rho_\star, \mathrm{rand}}$. We then draw a new value of the density as $\rho_\star \sim \mathcal{N}(\rho_{\star,\mathrm{true}}, \rho_{\star,\mathrm{true}}\times f_{\rho_\star, \mathrm{rand}})$. It is important to note that the transit chord ratio method is not very sensitive to $\rho_\star$ because the process of using ratios of transit chord lengths allows this dependency to partially cancel out.

\begin{figure*}
\centering
\epsscale{0.9}
\plotone{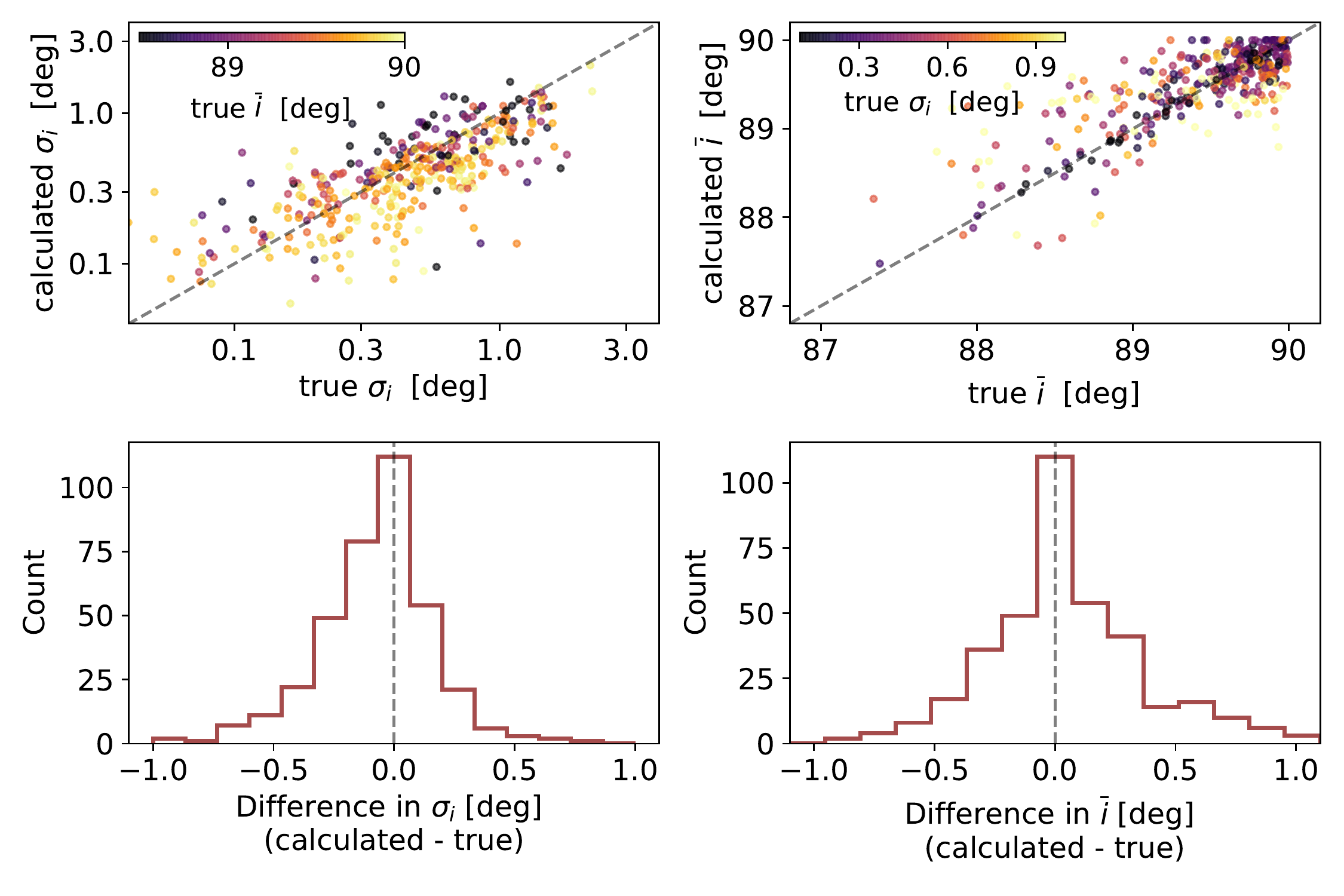}
\caption{Calculated values versus true values of the sky-plane inclination dispersion, $\sigma_i$, and mean inclination, $\overline{i}$ of 374 simulated SysSim systems with four or more detected planets. True values of $\overline{i}$ greater than $90^{\circ}$ are reported as $180^{\circ} - \overline{i}$, corresponding to a reflection across the $b=0$ chord. The top left panel shows the calculated values of $\sigma_i$ from the transit chord ratio method versus the true values, with the colorbar corresponding to the true mean inclination. The bottom left panel shows the distribution of differences in $\sigma_i$ values. The top right panel shows the calculated values of $\overline{i}$ versus the true values, with the colorbar corresponding to the true inclination dispersion. The bottom right panel shows the distribution of differences in $\overline{i}$.} 
\label{fig: True_vs_calculated_sigma_i_and_bar_i}
\end{figure*}

We use our method to estimate $\overline{i}$ and $\sigma_i$ in each SysSim 4+ system using $T$, $P$, and $\rho_\star$ values as inputs. Figure \ref{fig: True_vs_calculated_sigma_i_and_bar_i} shows the comparison between the derived values and true values of $\overline{i}$ and $\sigma_i$. There is good agreement overall, with relatively tight clustering of the calculated vs. true values around the one-to-one line. The median and 16th and 84th percentiles of the distribution of differences in $\sigma_i$ are ${-0.04^{\circ}}^{+0.16^{\circ}}_{-0.24^{\circ}}$. In addition, the offset of the calculated $\sigma_i$ values from the true values is weakly correlated with the true $\overline{i}$. This occurs because systems with $\overline{i}$ far from $90^{\circ}$ are more likely to have all planets transiting on one hemisphere of the star, which yields $\sigma_i$ values that are more accurate but biased high due to the effects of the unmodeled eccentricities. On the contrary, systems with $\overline{i}$ close to $90^{\circ}$ are more likely to have some planets transiting on separate hemispheres, which is not captured within the transit chord ratios and leads to underestimated $\sigma_i$ values. As for the $\overline{i}$ estimates, the median and 16th and 84th percentiles of the distribution of differences in $\overline{i}$ are ${0.03^{\circ}}^{+0.29^{\circ}}_{-0.28^{\circ}}$. We also compare (but do not plot) the true values of $\sigma_i$ with $\sigma_{i,\mathrm{inv}}$, the dispersion of the inclinations with respect to the invariable plane. We find that $\mathrm{median}(\sigma_{i,\mathrm{inv}}/\sigma_i)\approx 1.6$.

Before proceeding to utilize $\overline{i}$ and $\sigma_i$ to calculate transit probabilities, we introduce an additional scaling factor to $\sigma_i$. Recall that $\sigma_i$ is the sky-plane inclination dispersion of the observed \textit{transiting} planets. By definition, $\sigma_i$ does not contain information about the sky-plane inclinations of the \textit{non-transiting} planets. In order to approximately rectify this, we consider each SysSim system with four or more detected planets, and we calculate the sky-plane inclination dispersion, $\sigma_{i,\mathrm{all}}$, of \textit{all} planets in the system (i.e. including undetected planets) using the true values of the inclinations. We then compute the ratio of $\sigma_{i,\mathrm{all}}$ and the calculated $\sigma_i$ of the detected planets (as defined in equation \ref{eq: sigma_i}). We find that $\mathrm{median}(\sigma_{i,\mathrm{all}}/\sigma_i) = 1.2$. We thus define a scaled sky-plane inclination dispersion, $\sigma_i' = 1.2\sigma_i$, that now approximately reflects the spread of inclinations of the full system. We will use $\sigma_i$ whenever we're referring to the dispersion measured from the detected planets and $\sigma_i'$ whenever we're working in the context of transit probabilities, as in the next section.

\section{Model of Transit Probabilities and Detection Probabilities}
\label{sec: transit probabilities and detection probabilities}

\subsection{Transit probabilities}
\label{sec: transit probabilities}

Once the mean inclination, $\overline{i}$, and scaled inclination dispersion, $\sigma_i'$, have been estimated for a given system, it is possible to calculate the probability that a hypothetical planet with an arbitrary period transits the star. This requires an assumption that the inclinations of individual planets in the system are well-described by some distribution with mean and standard deviation equal to $\overline{i}$ and $\sigma_i'$. While there are multiple distributions one could consider, we find that a normal distribution, $i \sim \mathcal{N}\left(\overline{i},\sigma_i'\right)$, is a good approximation. 

\begin{figure}
\centering
\epsscale{1.2}
\plotone{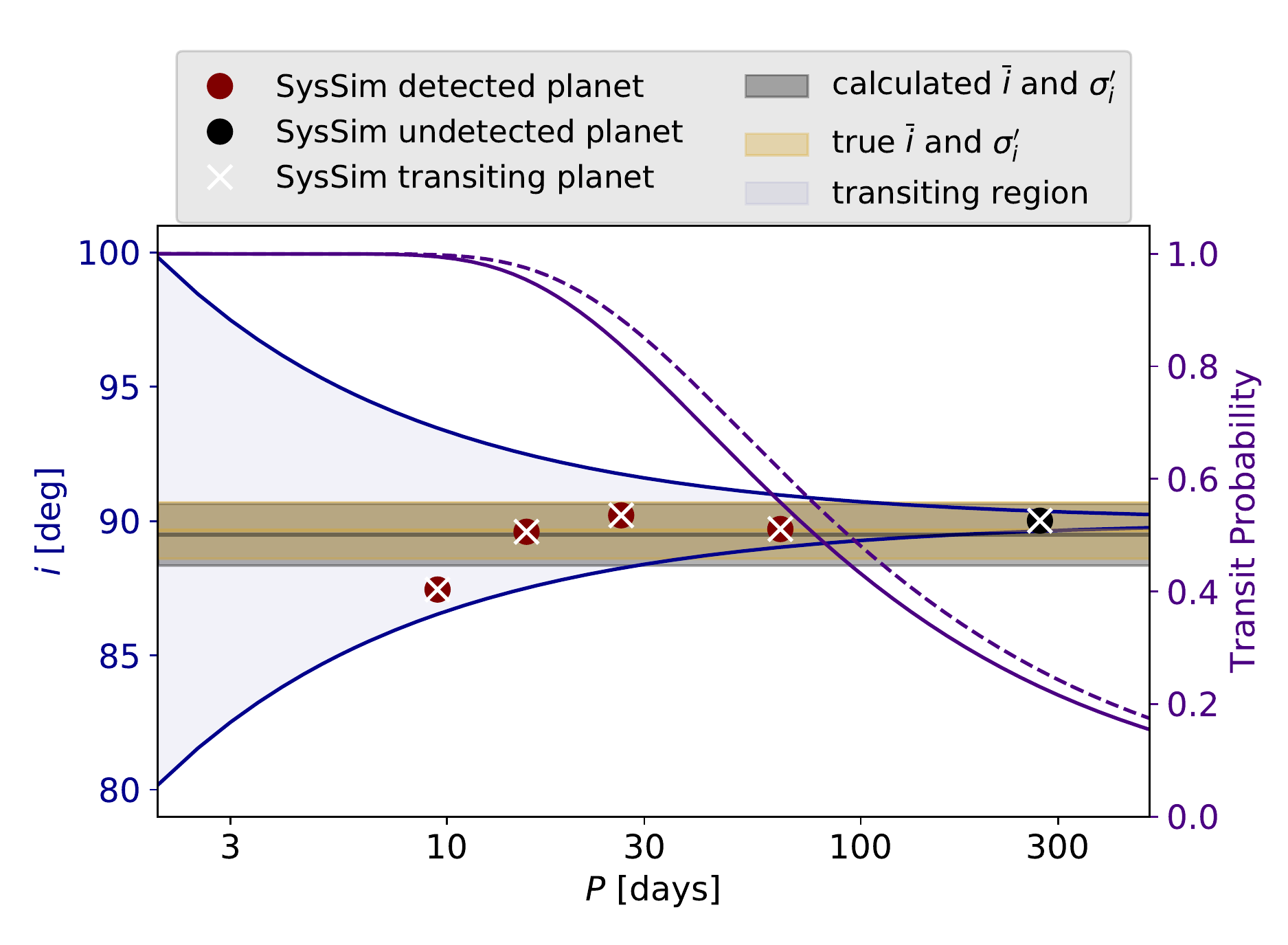}
\plotone{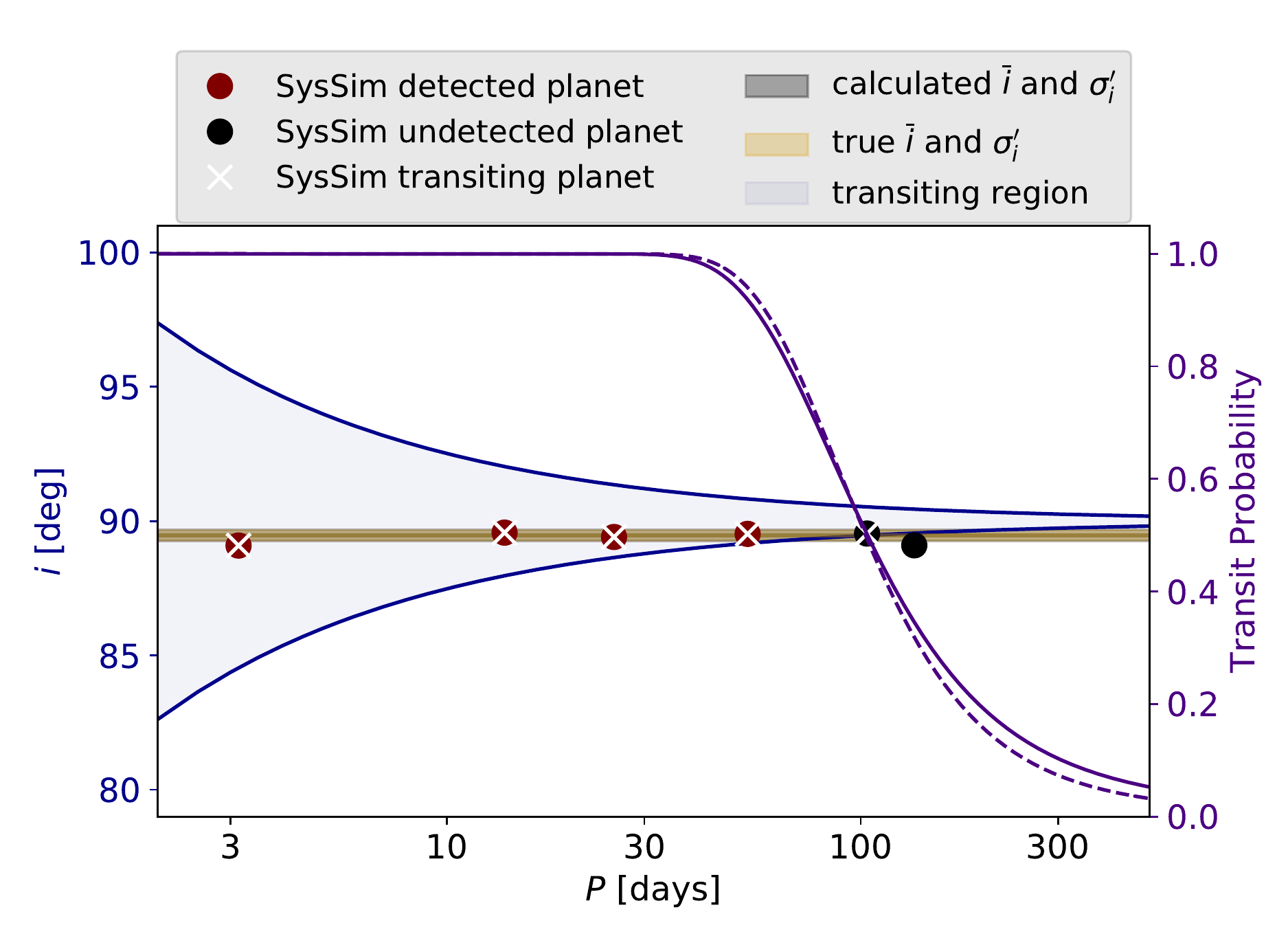}
\caption{Transit probability as a function of orbital period for two example SysSim simulated systems. The blue curves and solid envelope regions (left y-axis) show the inclinations at which a planet would transit with $|b| < b_{\mathrm{max}} = 1$ as a function of $P$ (equation \ref{eq: inclination to transit}). The gray (yellow) lines and banding indicate the calculated (true) values of $\overline{i}$ and $\sigma_i'$. The purple curves (right y-axis) show the transit probability, $p_{\mathrm{trans}}$, as a function of period (equation \ref{eq: transit probability}), with the solid (dashed) curve corresponding to the calculated (true) values of $\overline{i}$ and $\sigma_i'$. The maroon (black) dots indicate the detected (undetected) planets in the synthetic SysSim system. The white `x' points indicate the transiting planets in the system. } 
\label{fig: Transit_probability_example}
\end{figure}

The calculation of the transit probability first requires specifying the bounded range of inclinations for which a planet is transiting. Again assuming circular orbits, a planet transits with $|b| < b_{\mathrm{max}}$ if its inclination falls between ${i_\mathrm{min} < i < \pi - i_\mathrm{min}}$, where
\begin{align}
i_\mathrm{min} \equiv \arccos{\left[b_{\mathrm{max}}\left(\frac{3\pi}{G\rho_\star P^2}\right)^{1/3}\right]}.
\label{eq: inclination to transit}
\end{align}
The probability, $p_{\mathrm{trans}}$, that a planet with period $P$ transits with $|b| < b_{\mathrm{max}}$ is thus
\begin{equation}
p_{\mathrm{trans}} = \mathrm{Prob}(i_\mathrm{min} < i < \pi - i_\mathrm{min}); \ \  i \sim \mathcal{N}\left(\overline{i},\sigma_i'\right).
\label{eq: transit probability}
\end{equation}
We illustrate the transit probability calculations using two examples of SysSim systems with four detected planets. Figure \ref{fig: Transit_probability_example} shows the inclination range within which a planet would transit with $|b| < b_{\mathrm{max}}=1$, as well as the transit probability assuming that $i \sim \mathcal{N}\left(\overline{i},\sigma_i'\right)$. The transit probability is unity out to a period large enough such that the transiting region crosses over the region of higher probability according to $i \sim \mathcal{N}(\overline{i},\sigma_i')$. There is close agreement between the two transit probability curves associated with the calculated (solid purple) and true (dashed purple) values of $\overline{i}$ and $\sigma_i'$. As in these example SysSim systems, each system has its own unique curve of transit probability versus $P$ given the set of parameters $\rho_\star$, $\overline{i}$, and $\sigma_i'$. One can thus evaluate the transit probability of observed or hypothetical planets at arbitrary periods.

\subsection{Detection probabilities}
\label{sec: detection probabilities}

Even if a planet is transiting, it is not guaranteed to be detected. Here we outline the calculation of the detection probability, which depends on the statistical significance of the transit signal. The Kepler transiting planet search (TPS) pipeline required a planet to be transiting at least three times and to have a statistical significance exceeding a threshold of $7.1\sigma$ in order to be detected \citep{2020AJ....160..159C}. The statistical significance of the transit signal is computed with the multiple event statistic (MES), which cannot be specified exactly without applying the TPS pipeline to a given lightcurve but can be approximately computed using the one-sigma depth function (OSDF). The OSDF is a data product provided by Kepler DR25 \citep{2018ApJS..235...38T} that quantifies the transit signal that would be expected to result in an MES equal to unity for a given target star, orbital period, and transit duration, after averaging over the epoch of transit \citep{2017ksci.rept...14B}. Thus, the expected MES is simply 
\begin{equation}
\mathrm{MES} = \delta/\mathrm{OSDF}, 
\end{equation}
where $\delta$ is the transit depth. Kepler DR25 provided large tables of OSDF values for each target star as a function of 14 transit durations and $\sim10^4$ orbital periods; we use the downsampled versions of these tables provided by \cite{2019AJ....158..109H}. Given a target star and a planet's $P$ and $T$, we use bilinear interpolation to the star's OSDF table and extrapolate whenever $P$ or $T$ are beyond the range.  

We can now define a detection efficiency model. We build upon the model from \cite{2019AJ....158..109H}, which combines the probability of a transiting planet being detected and passing vetting, such that it is labeled a planet candidate. The model is calibrated to the Kepler DR25 pixel-level transit injection tests  \citep{2017ksci.rept...18C} and corresponding robovetter results \citep{2017ksci.rept...22C}. The probability of the planet passing through both detection and vetting is \citep{2019AJ....158..109H}\footnote{Note that there is a typo in this equation in \cite{2019AJ....158..109H}; there, the equation contains $\beta_{N_{\mathrm{tr}}} \times \mathrm{MES}$ in place of $\mathrm{MES}/\beta_{N_{\mathrm{tr}}}$.}
\begin{equation}
p_{\mathrm{det\&vet}}(\mathrm{MES},N_{\mathrm{tr}}) = c_{N_{\mathrm{tr}}}  \frac{\gamma(\alpha_{N_{\mathrm{tr}}}, \mathrm{MES}/\beta_{N_{\mathrm{tr}}})}{\Gamma(\alpha_{N_{\mathrm{tr}}})},
\label{eq: p_det_vet}
\end{equation}
where $N_{\mathrm{tr}}$ is the number of valid transits observed by Kepler, and $\alpha_{N_{\mathrm{tr}}}$, $\beta_{N_{\mathrm{tr}}}$, and $c_{N_{\mathrm{tr}}}$ are $N_{\mathrm{tr}}$-dependent parameters provided in Table \ref{tab: detection probability coefficients}. The number of transits can be approximated as ${N_{\mathrm{tr}} = \mathrm{floor}(t_{\mathrm{obs}} f_0/P)}$, where $t_{\mathrm{obs}}$ and $f_0$ are the target-specific data span and duty cycle. 

We apply additional modifications to our model to account for biases that can occur specifically in multi-planet systems, as shown by \citet{2019MNRAS.483.4479Z}. For example, the Kepler TPS pipeline masks each existing transit detection before searching for the next highest SNR candidate, reducing the available photometry for lower SNR candidate searches. These completeness issues are relevant since the theoretical planets considered in this study have the longest periods in their systems, with the fewest number of available transits. To ensure our estimates are conservative, we consider the case where these additional biases impact all the outermost planets. 

The completeness functions provided by \citet{2019MNRAS.483.4479Z} do not account for vetting as done by \citet{2019AJ....158..109H}. However, the multiplicity completeness will not be improved through the vetting process. To re-normalize the \citet{2019MNRAS.483.4479Z} results for vetting, we take the difference of the \citet{2019MNRAS.483.4479Z} multi and single planet completeness functions and subtract that from the \citet{2019AJ....158..109H} values. We then fit for the new parameters. The results of this manipulation are provided in Table \ref{tab: detection probability coefficients}. We acknowledge that a more accurate measure of multiplicity completeness would involve an injection/recovery test of multi-planet systems through the entirety of Kepler TPS pipeline, a procedure which is beyond the scope of this paper. Our estimates provide a first-order approximation, and the results in the following sections exceed any expected changes from a more refined completeness assessment.

\begin{table}[t!]
\caption{Parameters for $p_{\mathrm{det\&vet}}$ (equation \ref{eq: p_det_vet}). We show the original values from \cite{2019AJ....158..109H} and our updated values, after accounting for the effects of planet multiplicity.}
\begin{tabular}{c c c c}
\hline
\hline
$N_{\mathrm{tr}}$ & $\alpha_{N_{\mathrm{tr}}}$ & $\beta_{N_{\mathrm{tr}}}$ & $c_{N_{\mathrm{tr}}}$ \\ 
\hline
\multicolumn{4}{c}{\cite{2019AJ....158..109H} values} \\ 
\hline
3 & 33.3884 & 0.264472 & 0.699093  \\
4 & 32.8860 & 0.269577 & 0.768366  \\
5 & 31.5196 & 0.282741 & 0.833673  \\
6 & 30.9919 & 0.286979 & 0.859865  \\
$7-9$ & 30.1906 & 0.294688 & 0.875042  \\
$10-18$ & 31.6342 & 0.279425 & 0.886144  \\
$19-36$ & 32.6448 & 0.268898 & 0.889724  \\
$\geq 37$ & 27.8185 & 0.32432 & 0.945075  \\
\\
\hline
\multicolumn{4}{c}{New values} \\ 
\hline
3 & 21.6423 & 0.493516 & 0.535406 \\
4 & 30.4538 &  0.317441 &  0.713207  \\
5 & 29.7524  &  0.324572 & 0.778662  \\
6 & 28.8620 & 0.333193 & 0.804942  \\
$7-9$ & 28.1473 & 0.341453 & 0.820238    \\
$10-18$ & 28.4430 &  0.335102 & 0.831143  \\
$19-36$ & 28.3006 & 0.334147  & 0.834582  \\
$\geq 37$ & 27.4994 & 0.352761 & 0.890567 \\
\hline
\label{tab: detection probability coefficients}
\end{tabular}
\end{table}

Before returning the final probability of detecting a planet, the probability in equation \ref{eq: p_det_vet} is multiplied by an additional factor called the window function (WF), or the fraction of unique transit ephemeris epochs that permit three or more transits to be observed as a function of orbital period \citep{2017ksci.rept...14B}. This factor is necessary and complex because it provides the probability that a planet has $N_{\mathrm{tr}} \geq 3$ given existing data gaps within the limited span of Kepler photometry, whereas equation \ref{eq: p_det_vet} assumes $N_{\mathrm{tr}} \geq 3$ to be the case. Similar to the OSDFs, Kepler DR25 provided tabulations of WFs for each star as a function of $T$ and $P$. We use the downsampled versions provided by \cite{2019AJ....158..109H}. We use bilinear interpolation to calculate the target star's WF as a function of $P$ and $T$, extrapolating whenever these values are beyond the range.  

\section{Validation Tests with Simulated Planetary Systems}
\label{sec: Validation Tests with SysSim}

\begin{figure*}
\centering
\epsscale{1.18}
\plotone{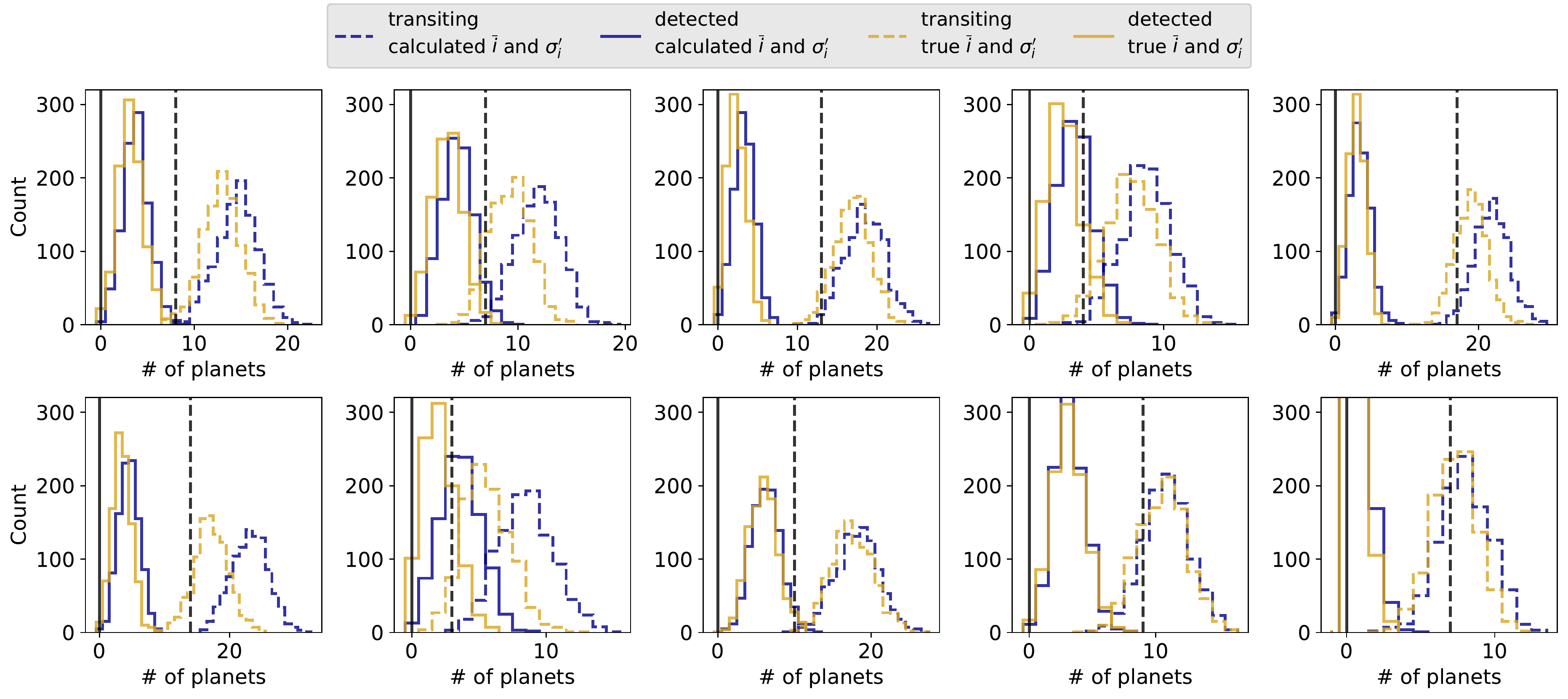}
\caption{Distributions of the number of next outer undetected planets that we would expect to be transiting, $N_{\mathrm{trans}}$, and detected, $N_{\mathrm{detect}}$, in systems with four or more detected planets within ten SysSim catalogs. Each panel represents a different catalog. The dashed histograms represent $N_{\mathrm{trans}}$, while the solid histograms represent $N_{\mathrm{detect}}$. The blue histograms use calculated values of $\overline{i}$ and $\sigma_i'$ (from the transit chord ratio method) for the $p_{\mathrm{trans}}$ calculation, while the yellow histograms values use true values of $\overline{i}$ and $\sigma_i'$. The dashed (solid) vertical black lines correspond to the true numbers of next outer undetected planets that are transiting (detected) in the catalog. } 
\label{fig: SysSim_next_outer_planet_distributions}
\end{figure*}

The previous section outlined the ingredients necessary to calculate an expected distribution of detections of outer planets orbiting beyond observed planets in multi-planet systems. Before applying this to the Kepler compact multis (Section \ref{sec: hypothetical planet experiments}), we will first test our methods on sub-populations of simulated planets from the SysSim maximum AMD model \citep{2020AJ....160..276H}. We will essentially ``re-simulate'' the transit and detection observations using the methods developed in Section \ref{sec: transit probabilities and detection probabilities} as a means of validating that our approach results in accurate total numbers of transiting and detected planets. We will specifically examine outer planets in SysSim systems with four or more detected planets (``4+ systems'') and verify that our calculations are in line with the true properties of the synthetic systems.

We begin with the same ten pairs of SysSim physical catalogs and observed catalogs as used in Section \ref{sec: transit chord ratio method validation using SysSim}. Recall that a single physical/observed catalog pair represents an entire Kepler population of planetary systems orbiting FGK dwarfs. We consider the 4+ systems only. In each system, we identify the innermost undetected planet beyond the detected planets, which we will call the ``next outer undetected planet''. For instance, in each example system in Figure \ref{fig: Transit_probability_example}, the fifth planet from left is the next outer undetected planet. Not all systems have such a planet; in $\sim25\%$ of systems, the outermost planet is a detected planet, so the next outer undetected planet is undefined. We focus on the remaining $\sim75\%$ of systems for which this planet is defined. The true number of detections in this sub-sample is zero by definition, so when we calculate the expected number of detections, we should get a result that is consistent with zero. 

For each physical/observed catalog pair, we calculate the distribution of the number of next outer undetected planets that we would expect to be transiting and detectable. We use the following steps: 
\begin{enumerate}
\item For each next outer undetected planet, we estimate the transit probability, $p_{\mathrm{trans}}$, according to equation \ref{eq: transit probability}, where $\overline{i}$ and $\sigma_i'$ are either the true values or the calculated values from the transit chord ratio method.
\item We approximate each planet's transit duration and transit depth (if it were to transit) using scalings to the duration and depth of the outermost observed planet, 
\begin{equation}
\begin{split}
T_{\mathrm{next}} &= T_{\mathrm{out}}(P_{\mathrm{next}}/P_{\mathrm{out}})^{1/3} \\
\delta_{\mathrm{next}} &= \delta_{\mathrm{out}}(R_{p,\mathrm{next}}/R_{p,\mathrm{out}})^2.
\end{split}
\end{equation}
\item Using the period, $P_{\mathrm{next}}$, transit duration, $T_{\mathrm{next}}$, and transit depth, $\delta_{\mathrm{next}}$, we estimate the OSDF, WF, and MES for each planet using the target-specific data tables and the procedure described in Section \ref{sec: detection probabilities}. 
\item  We use the MES and WF to estimate the detection probability, $p_{\mathrm{det\&vet}}$, using equation \ref{eq: p_det_vet}. 

\item Using the calculated probabilities, we draw two Bernoulli random variables, 
\begin{equation}
\begin{split}
X_{\mathrm{trans}} &\sim \mathrm{Bernoulli}(p_{\mathrm{trans}}) \\ 
X_{\mathrm{detect}} &\sim \mathrm{Bernoulli}(p_{\mathrm{trans}} \times p_{\mathrm{det\&vet}}),
\end{split}
\end{equation}
with the $X_{\mathrm{trans}}$ indicating whether the planet is transiting and $X_{\mathrm{detect}}$ indicating whether the planet is both transiting and detected. 

\item The sum of Bernoulli random variables across all the systems in a given catalog pair gives a number of ``successes'' (transits or detections),
\begin{equation}
\begin{split}
N_{\mathrm{trans}} &= \sum_{N_{4+}}X_{\mathrm{trans}} \\ 
N_{\mathrm{detect}} &= \sum_{N_{4+}}X_{\mathrm{detect}}.
\end{split}
\end{equation}

\item Finally, by repeating steps five and six 1000 times, we create distributions of $N_{\mathrm{trans}}$ and $N_{\mathrm{detect}}$ for each catalog pair (with 1000 values in each distribution).
\end{enumerate}

Figure \ref{fig: SysSim_next_outer_planet_distributions} shows the distributions of $N_{\mathrm{trans}}$ and $N_{\mathrm{detect}}$ of the next outer undetected planets across ten SysSim catalog pairs. The first thing to note is that the distributions of $N_{\mathrm{trans}}$ and $N_{\mathrm{detect}}$, which use calculated values of $\overline{i}$ and $\sigma_i'$ for the $p_{\mathrm{trans}}$ calculation (blue histograms), agree well with the corresponding distributions that use the true values of $\overline{i}$ and $\sigma_i'$ (yellow histograms). Thus, we can trust that using calculated values of these key parameters will provide accurate distributions when we proceed to working with the Kepler multis.

A second observation from Figure \ref{fig: SysSim_next_outer_planet_distributions} is that the distributions of $N_{\mathrm{trans}}$ and $N_{\mathrm{detect}}$ are in reasonable agreement with the true values of $N_{\mathrm{trans}}$ and $N_{\mathrm{detect}}$. (Recall that the true value of $N_{\mathrm{detect}}$ is always zero because these experiments are conditioned upon predictions for the first outer \textit{undetected} planet in each system.) The distributions are consistent with the true values to within $\sim1-3\sigma$, with the average being $\sim2.2\sigma$ for $N_{\mathrm{trans}}$ and $\sim2.4\sigma$ for $N_{\mathrm{detect}}$. However, the distributions are always biased high. This is by construction of our exploration. Because the next outer undetected planets were missed within the actual SysSim simulations, they are more likely to be non-transiting than other planets in the same system. Thus, our assumption of $i \sim \mathcal{N}(\overline{i}, \sigma_i')$ in $p_{\mathrm{trans}}$ (equation \ref{eq: transit probability}) partially breaks down and overestimates the transit probability. Regardless, the fact that the distributions agree with the true $N_{\mathrm{trans}}$ and $N_{\mathrm{detect}}$ to within $\lesssim3\sigma$ ensures that we can reliably use these methods on the observed Kepler multis, as long as any results we find are significant by much more than $3\sigma$. 

%Before proceeding to the observed systems, it is helpful to discuss the dominant origins of non-detections among transiting planets in the simulated systems. In most cases, such planets are undetected because of too low MES, rather than too few transits. Roughly $\sim50\%$ of the next outer undetected planets have $\mathrm{MES}<7$ (either due to small radius or long period), whereas less than $1\%$ of them have three or fewer transits. However, a sizeable $\sim8\%$ have only four transits, so some non-detections can indeed arise due to low transit number.     

\section{Detection Expectations for Hypothetical Outer Planets in Kepler Multi-Planet Systems}
\label{sec: hypothetical planet experiments}

In the previous section, we used populations of simulated planetary systems to validate our calculated expectations of the number of planets that would be transiting and detected. We now extend these methods to Kepler high-multiplicity systems. We aim to answer the questions: If the Kepler multis hosted additional outer planets with similar properties as the known planets, how many of them would we expect to be transiting and detectable? Moreover, are these expectations reconcilable with the lack of detections beyond the outermost known transiting planets?

\subsection{Sample of Kepler multi-planet systems}
\label{sec: sample selection}

We begin by defining our sample of Kepler multi-planet systems. We use the Kepler DR25 KOI catalog \citep{2018ApJS..235...38T, koidr25} as our starting point, using all planets with ``confirmed'' and ``candidate'' dispositions. Where possible, we replace the stellar parameters and planet radii in the DR25 catalog with parameters from the Gaia-Kepler Stellar Properties Catalog \citep{2020AJ....160..108B, 2020AJ....159..280B}. In addition, we apply a small set of quality cuts. We consider only planets smaller than $16 \ R_{\oplus}$ with fractional radius uncertainties less than 100\%. To avoid stars with large systematic radius errors, we discard targets for which \cite{2017AJ....153...71F} found a companion star that contributed more than 5\% of the light in the photometric aperture. After these cuts, we are left with 64 Kepler systems with four or more observed transiting planets (``4+ systems'').

\begin{figure}
\centering
\epsscale{1}
\plotone{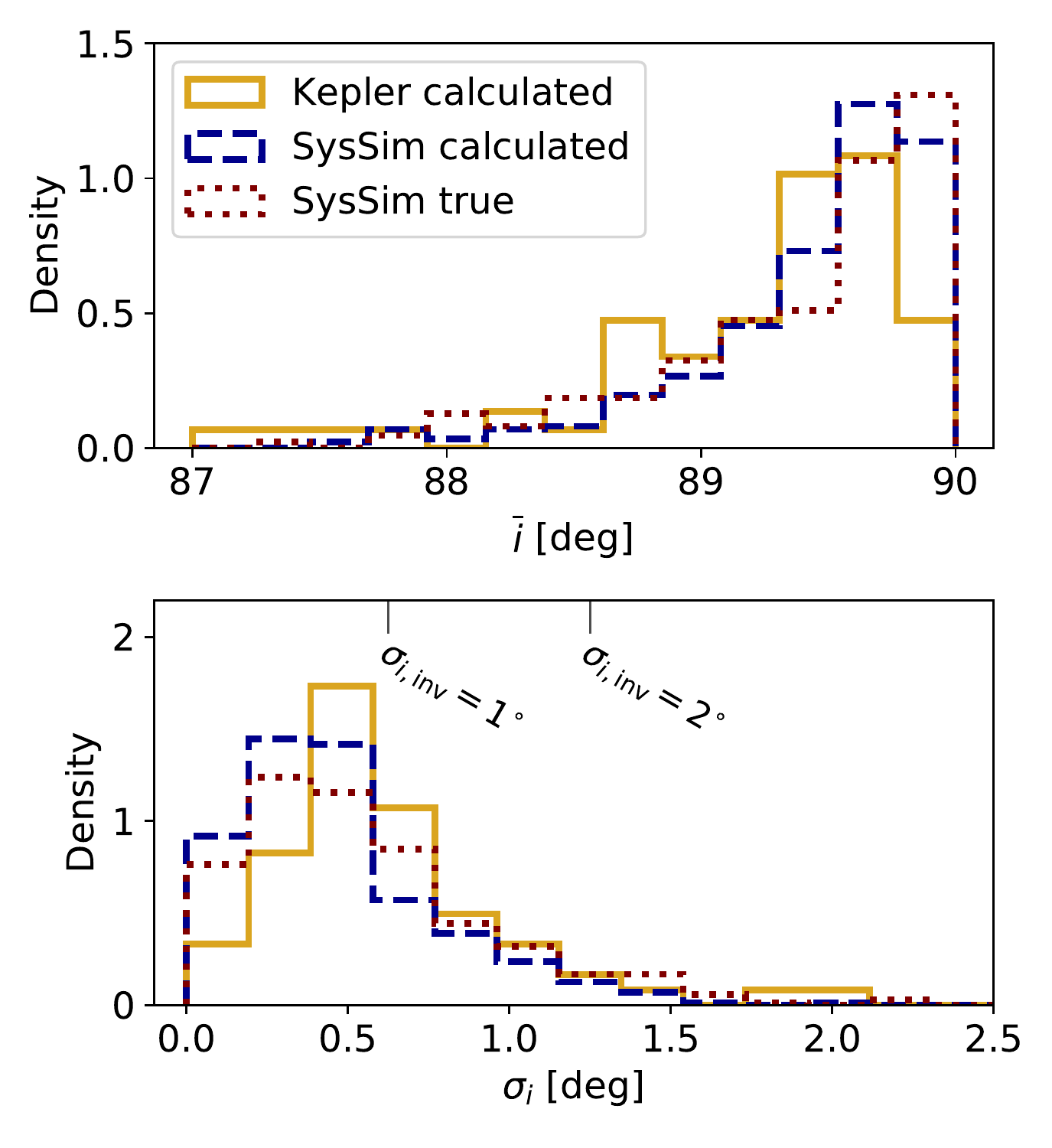}
\caption{Comparison of $\overline{i}$ and $\sigma_i$ distributions for Kepler 4+ systems and SysSim 4+ systems. The top (bottom) panels show the distributions of $\overline{i}$ ($\sigma_i$), with the calculated values for Kepler multis in solid yellow, the calculated values for SysSim multis in dashed blue, and the true values for SysSim multis in dotted maroon. True values of $\overline{i}$ greater than $90^{\circ}$ are reported as $180^{\circ} - \overline{i}$, corresponding to a reflection across the $b=0$ chord. Benchmark values of the dispersion of the inclinations with respect to the invariable plane, $\sigma_{i,\mathrm{inv}}\approx1.6\sigma_i$ (Section \ref{sec: transit chord ratio method validation using SysSim}), are also labeled.} 
\label{fig: Kepler_vs_SysSim_sigma_i_and_i_bar}
\end{figure}

We now apply the transit chord ratio method (Section \ref{sec: transit chord ratio method}) to the transiting planets in the Kepler 4+ systems to estimate the mean sky-plane inclination, $\overline{i}$, and the sky-plane inclination dispersion, $\sigma_i$ (from which the scaled dispersion $\sigma_i' = 1.2\sigma_i$ is subsequently derived). Figure \ref{fig: Kepler_vs_SysSim_sigma_i_and_i_bar} shows the distributions of $\overline{i}$ and $\sigma_i$ for the Kepler 4+ systems and the SysSim 4+ systems examined in Sections \ref{sec: transit chord ratio method validation using SysSim} and \ref{sec: Validation Tests with SysSim}. For the SysSim data, we show both the calculated values from the transit chord ratio method and the true values. Overall, the distributions for the Kepler systems bear close resemblance to those for the SysSim systems. This offers another validation of the $\overline{i}$ and $\sigma_i$ estimates, which will be used in transit probability calculations in the next section.

\subsection{Set-up of hypothetical planet experiments}

We now construct several experiments consisting of ``hypothetical planets'' orbiting beyond the observed planets in Kepler 4+ systems. In general, the hypothetical planets are designed to emulate the continuation of the ``peas-in-a-pod'' architectures \citep{2018AJ....155...48W, 2017ApJ...849L..33M} by positing the existence of one additional outer planet in each system. We first assign the periods, $P_{\mathrm{hypo}}$, and radii, $R_{p,\mathrm{hypo}}$, of the hypothetical planets. We consider three separate approaches, outlined below and denoted ``fixed'', ``random'', and ``targeted'' sampling. 

\begin{figure*}
\centering
\epsscale{0.8}
\plotone{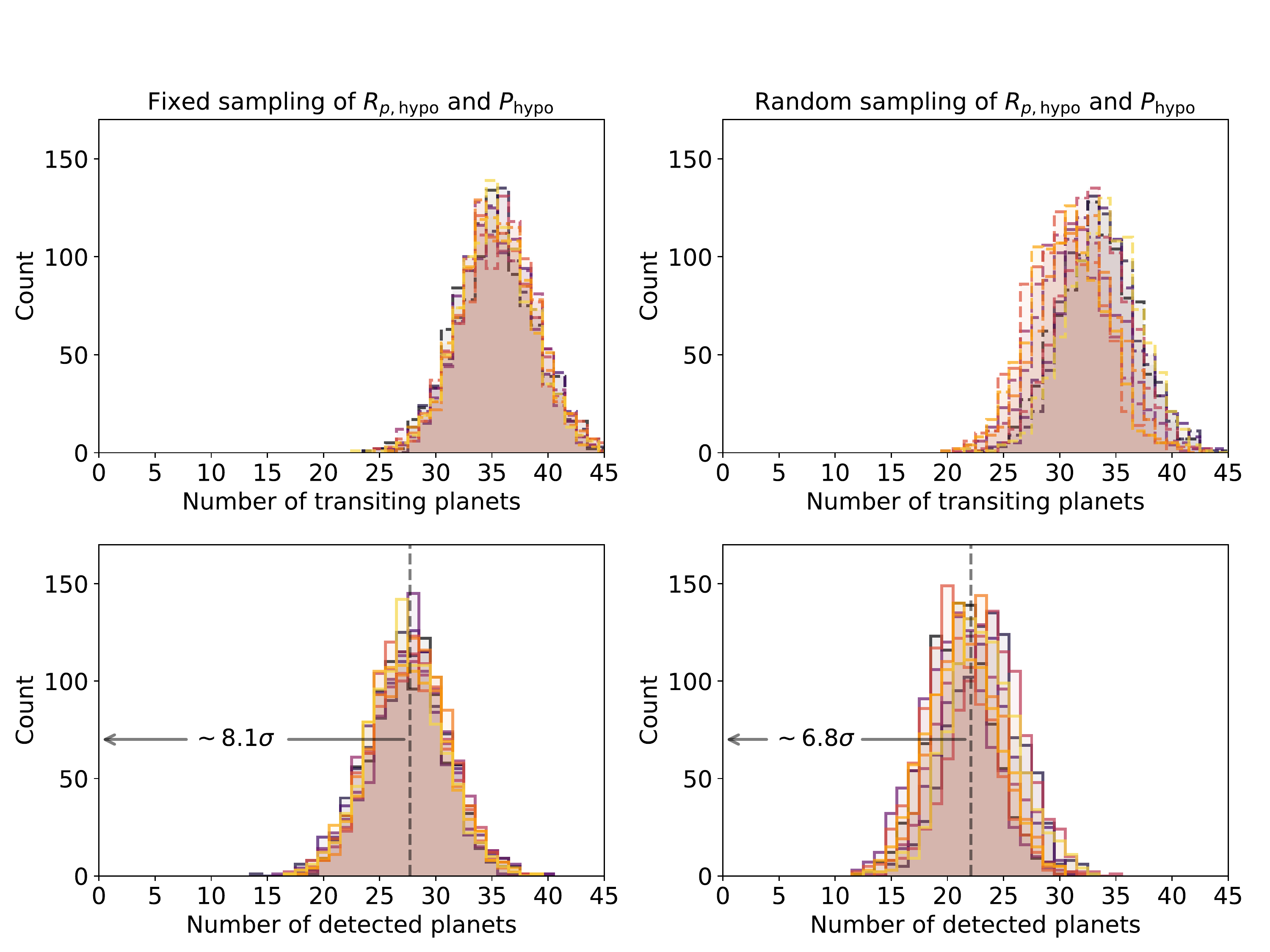}
\caption{Distributions of the number of ``hypothetical planets'' that we would expect to be transiting, $N_{\mathrm{trans}}$, and detected, $N_{\mathrm{detect}}$, if they existed at the outer edges of the 64 Kepler systems with four or more known transiting planets. The left column corresponds to the fixed sampling approach with $R_{p,\mathrm{hypo}} = R_{p,\mathrm{out}}$ and
$P_{\mathrm{hypo}}/P_{\mathrm{out}} = P_{\mathrm{out}}/P_{\mathrm{2nd-out}}$. The right column corresponds to the random sampling approach with $R_{p,\mathrm{hypo}}/R_{p,\mathrm{out}}$ and $(P_{\mathrm{hypo}}/P_{\mathrm{out}})/(P_{\mathrm{out}}/P_{\mathrm{2nd-out}})$ sampled from empirical distributions. The dashed histograms in the top panels represent $N_{\mathrm{trans}}$, while the solid histograms in the bottom panels represent $N_{\mathrm{detect}}$. The different colors represent ten separate iterations and have no significance otherwise. } 
\label{fig: fixed sampling and random sampling results}
\end{figure*}

\textit{Fixed sampling:} The most straightforward approach in assigning $P_{\mathrm{hypo}}$ and $R_{p,\mathrm{hypo}}$ is to consider that they perfectly follow the ``peas-in-a-pod'' patterns \citep{2018AJ....155...48W, 2017ApJ...849L..33M}, or the statistical tendency for systems to exhibit intra-system uniformity in planetary radii and period ratios. Accordingly, we can assign
\begin{equation}
\begin{split}
&R_{p,\mathrm{hypo}} = R_{p,\mathrm{out}} \\
&P_{\mathrm{hypo}}/P_{\mathrm{out}} = P_{\mathrm{out}}/P_{\mathrm{2nd-out}},
\label{eq: fixed sampling}
\end{split}
\end{equation}
where $P_{\mathrm{out}}$ and $R_{p,\mathrm{out}}$ are the period and radius of the outermost known planet, and $P_{\mathrm{2nd-out}}$ is the period of the second outermost known planet. (This is identical to the initial experiments shown in Section \ref{sec: heuristic calculation}.) 

\textit{Random sampling:} The patterns of intra-system uniformity of radii and period ratios have an inherent degree of scatter \citep[e.g.][]{2021ApJ...920L..34M}. Thus, a second and more physical approach to assigning $P_{\mathrm{hypo}}$ and $R_{p,\mathrm{hypo}}$ is to utilize empirical distributions of radius ratios and ratios of period ratios from our sample. We assign 
\begin{equation}
\begin{split}
&R_{p,\mathrm{hypo}}/R_{p,\mathrm{out}} = R_{p,i+1}/R_{p,i}, \\
&P_{\mathrm{hypo}}/P_{\mathrm{out}} = P_{\mathrm{out}}/P_{\mathrm{2nd-out}}\left(\frac{P_{i+1}/P_i}{P_{i}/P_{i-1}}\right),
\label{eq: random sampling}
\end{split}
\end{equation}
where $R_{p,i+1}/R_{p,i}$ is randomly drawn from the observed distribution of radius ratios of adjacent planets, and $(P_{i+1}/P_i)/(P_{i}/P_{i-1})$ is drawn from the observed distribution of ratios of period ratios of adjacent planet pairs. However, in an effort to preserve covariances of the distributions with respect to other variables (e.g. correlations between $R_{p,i+1}/R_{p,i}$ and $P_{i+1}/P_i$), our random draws are not taken from the distributions as a whole but rather from pre-defined subsets. We bin the distribution of $R_{p,i+1}/R_{p,i}$ in five bins of $P_i$ and four bins of $P_{i+1}/P_i$ with an average of about 40 points in each bin. The bins are chosen so as to capture the large-scale correlations within the distribution, and the overall results are not sensitive to them. When sampling from the distribution of $R_{p,i+1}/R_{p,i}$, we identify the appropriate bin and only sample from the corresponding subset of the distribution. Similarly, we bin the distribution of $(P_{i+1}/P_i)/(P_{i}/P_{i-1})$ in six bins of the inner period ratio, $P_{i}/P_{i-1}$, and we sample from the distribution by first identifying the appropriate bin and only drawing from the corresponding subset of the distribution. 

\textit{Targeted sampling:} Finally, a third approach of assigning $P_{\mathrm{hypo}}$ and $R_{p,\mathrm{hypo}}$ is designed to allow us to systematically vary these parameters and observe the resulting changes to the distributions of $N_{\mathrm{trans}}$ and $N_{\mathrm{detect}}$. Specifically, we sample $R_{p,\mathrm{hypo}}/R_{p,\mathrm{out}} \in [0.2, 1]$ and $(P_{\mathrm{hypo}}/P_{\mathrm{out}})/(P_{\mathrm{out}}/P_{\mathrm{2nd-out}}) \in [1,4]$, thus probing configurations in which the hypothetical planet is either smaller or further away than the ``peas-in-a-pod'' expectation. We target $P_{\mathrm{hypo}}$ and $R_{p,\mathrm{hypo}}$ one at a time using this sampling method, with the other parameter assigned using the ``fixed sampling'' approach.

In addition to the targeted sampling of $P_{\mathrm{hypo}}$ and $R_{p,\mathrm{hypo}}$, we also extend our targeted sampling experiments to two other parameters. First, we consider assigning hypothetical planets to only a fraction of systems, rather than all systems. We sample this fraction in the range $f_{\mathrm{hypo}} \in [0.1, 1]$. This allows us to investigate how many systems can have additional outer planets and still be consistent with the observations. Lastly, in order to explore the effects of potentially biased estimations of the sky-plane inclination dispersion, we consider a constant scaling factor, $\gamma_{\sigma_i'}$, such that $\sigma_{i,\mathrm{new}}'=\gamma_{\sigma_i'}\sigma_i'$. We sample the scaling factor in the range $\gamma_{\sigma_i'} \in [1,10]$. We reiterate that we use targeted sampling on only one of the four parameters ($P_{\mathrm{hypo}}$, $R_{p,\mathrm{hypo}}$, $f_{\mathrm{hypo}}$, and $\gamma_{\sigma_i'}$) at a time, with all others assigned using the ``fixed sampling'' approach (which, in the case of the latter two parameters, means $f_{\mathrm{hypo}}=1$ and $\gamma_{\sigma_i'}=1$).

With the sampling approach defined and $P_{\mathrm{hypo}}$ and $R_{p,\mathrm{hypo}}$ assigned for each hypothetical planet, we proceed to calculate the distribution of the number of these planets that we would expect to be transiting and detectable. We use the same seven steps as outlined in Section \ref{sec: Validation Tests with SysSim}, where the transit probabilities and detection probabilities are calculated for each hypothetical planet, and repeated trials of summed Bernoulli random variables yield distributions of $N_{\mathrm{trans}}$ and $N_{\mathrm{detect}}$.

\subsection{Results from fixed sampling and random sampling}

Figure \ref{fig: fixed sampling and random sampling results} shows the results of the fixed sampling and random sampling approaches. Ten iterations of the distributions are shown, indicating the degree of variability resulting from the sampling of $P_{\mathrm{hypo}}$ and $R_{p,\mathrm{hypo}}$ combined with the seven steps outlined in Section \ref{sec: Validation Tests with SysSim}. Immediately, we can see that the distributions of $N_{\mathrm{detect}}$ are inconsistent with zero detections in both the fixed sampling and random sampling approaches. With the fixed sampling approach, the mean and standard deviation of the composite distribution of $N_{\mathrm{detect}}$ (consisting of the ten separate distributions pooled together) is $27.55 \pm 3.39$ detected planets, or $\sim8.1\sigma$ greater than zero detections. This indicates that $\sim44\%$ of the 64 Kepler 4+ systems would be expected to show detections of the hypothetical outer planets, if they existed.

As for the distributions resulting from the random sampling approach, there is more variability between different iterations due to each iteration having a distinct set of $P_{\mathrm{hypo}}$ and $R_{p,\mathrm{hypo}}$ values. The median and standard deviation of the composite distribution of $N_{\mathrm{detect}}$ is $22.11 \pm 3.25$ detected planets, or $\sim6.8\sigma$ greater than zero detections. This indicates that the hypothetical planets would be detectable in $\sim35\%$ of the 64 Kepler 4+ systems.

Both the fixed sampling and random sampling approaches indicate the same general result: If we posit the existence of additional planets orbiting beyond the known transiting planets with properties dictated by the expected ``peas-in-a-pod'' architectures, then these planets would be detectable in roughly $\sim 35\%-45\%$ of systems, yielding a $\sim 7-8\sigma$ discrepancy with the lack of detections of such planets in the observed systems. However, it is important to note that the distributions of $N_{\mathrm{detect}}$ are always positive by definition, so the more meaningful comparison is that with the corresponding results for the SysSim simulated systems (Section \ref{sec: Validation Tests with SysSim}; Figure \ref{fig: SysSim_next_outer_planet_distributions}), where we found that the $N_{\mathrm{detect}}$ distributions were biased high but always consistent with zero to within $3\sigma$.  

A final point is that this result is a lower limit on the expected number of detections since we only assigned one hypothetical planet per system. If we considered more than one outer planet per system, this would potentially generate an even larger number of expected detections. For instance, one possible experiment would be to add multiple sequential hypothetical planets to each system, assigned as $P_{\mathrm{hypo},i} = P_{\mathrm{out}}(P_{\mathrm{out}}/P_{\mathrm{2nd-out}})^i$ and restricted to $P_{\mathrm{hypo},i} < 500$ days. We find that this would involve a total of 327 hypothetical planets, compared to the 64 in this study. 

\subsection{Results from targeted sampling}
\label{sec: Results from targeted sampling}

\begin{figure*}
\centering
\epsscale{1.2}
\plotone{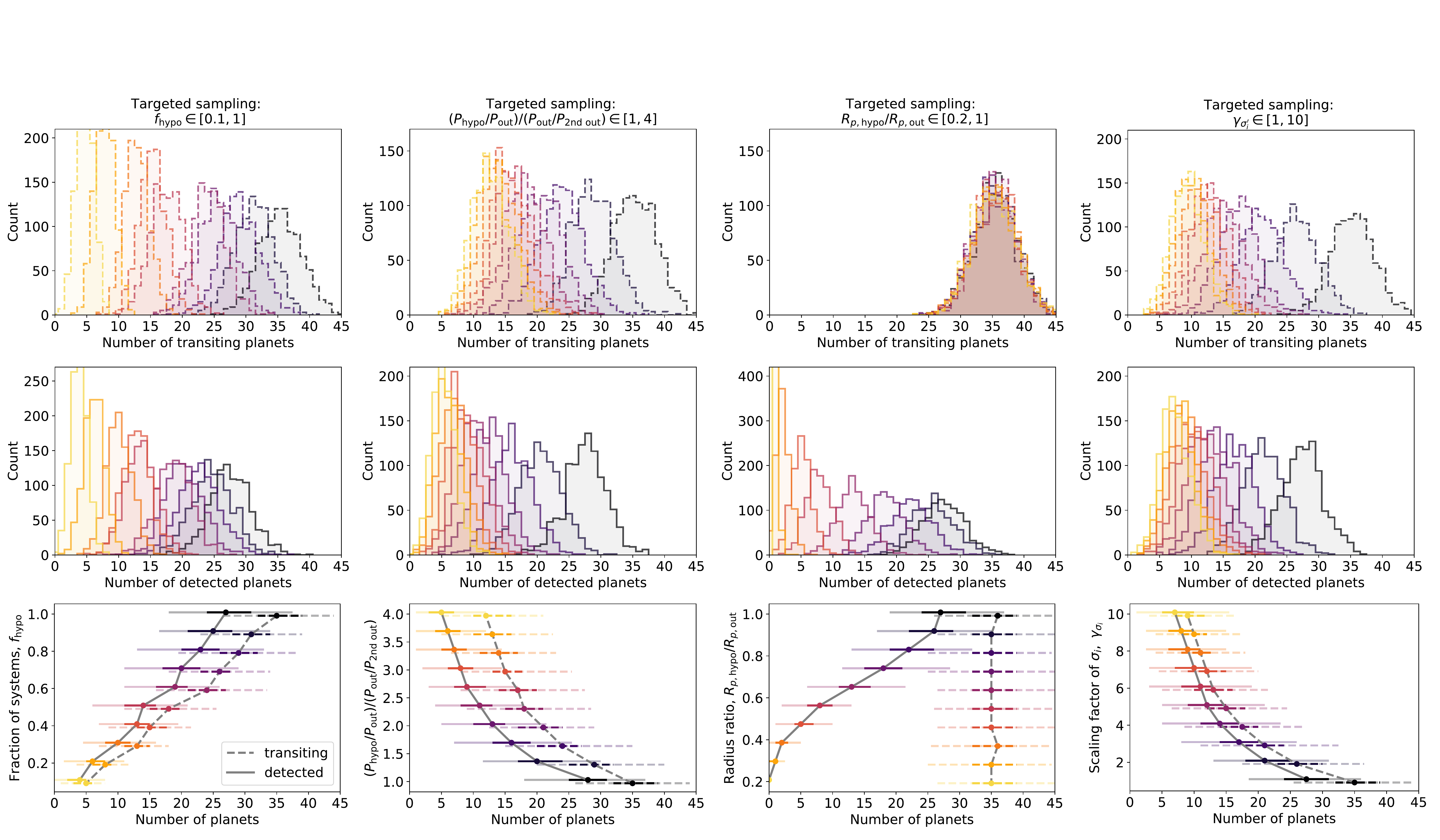}
\caption{Distributions of the number of ``hypothetical planets'' that we would expect to be transiting, $N_{\mathrm{trans}}$, and detected, $N_{\mathrm{detect}}$, if they existed at the outer edges of the 64 Kepler systems with four or more known transiting planets. All columns indicate the results of the targeted sampling approach, with the columns corresponding to (from left to right) sampling of $f_{\mathrm{hypo}}$ (the fraction of systems with hypothetical planets), $(P_{\mathrm{hypo}}/P_{\mathrm{out}})/(P_{\mathrm{out}}/P_{\mathrm{2nd-out}})$, $R_{p,\mathrm{hypo}}/R_{p,\mathrm{out}}$, and $\gamma_{\sigma_i'}$ (the scaling factor of the inclination dispersion). These parameters are sampled from 10 evenly-spaced values within the ranges indicated at the top of the columns. The dashed histograms in the top panels represent $N_{\mathrm{trans}}$, and the solid histograms in the middle panels represent $N_{\mathrm{detect}}$. The colors indicate the value of the targeted parameter, as shown in the bottom panels, where the $1\sigma$ and $3\sigma$ intervals around the medians of the distributions are indicated with the darker and lighter errorbars. The two curves in the bottom panels are artificially offset vertically such that the errorbars can be seen clearly.} 
\label{fig: targeted sampling results}
\end{figure*}

The discrepancy between the expected number of detections and the lack of such detections in the real systems indicates that one or more of our assumptions are breaking down. These assumptions and their related implications are enumerated below: 
\begin{enumerate}
\item We assumed that each system has a single additional planet orbiting beyond the known transiting planets. However, perhaps not all systems have a such a planet. This would imply that the architectures of compact multis are truncated at a detectable orbital period in at least some fraction of systems.   
\item We assumed that $P_{\mathrm{hypo}}/P_{\mathrm{out}} \approx P_{\mathrm{out}}/P_{\mathrm{2nd-out}}$. However, perhaps the period ratios are on average larger at sufficiently large orbital period.\footnote{We searched for a trend between orbital period and period ratio in the observed sample of Kepler 4+ systems and found none. However, it is still possible that the period ratios increase at orbital periods at or near the edge of detectability.}
\item We also assumed that $R_{p,\mathrm{hypo}} \approx R_{p,\mathrm{out}}$. However, perhaps the radii of planets are on average smaller at sufficiently large orbital period.
\item Finally, we also assumed a mutual inclination dispersion for each system based on our estimation from the transit chord ratio method. Although we validated the method in Section \ref{sec: transit chord ratio method}, it is possible that the $\sigma_i'$ estimates are systematically low or that the true dispersion increases at large periods. 
\end{enumerate}

The targeted sampling approach described earlier in this section allows us to investigate these four possibilities by systematically varying $f_{\mathrm{hypo}}$, $P_{\mathrm{hypo}}$, $R_{p,\mathrm{hypo}}$, and $\gamma_{\sigma_i'}$, which directly map to the four possibilities outlined above. All other parameters are held fixed while the targeted parameters are varied. Figure \ref{fig: targeted sampling results} shows the results of the targeted sampling. The bottom panels summarize the distributions by showing the medians and $1\sigma$ and $3\sigma$ intervals as a function of the targeted parameter. This allows us to visualize how much the targeted parameter must be varied in order to have the $N_{\mathrm{detect}}$ distribution be approximately consistent with zero detections to within $3\sigma$.

As shown in the figure, the $N_{\mathrm{detect}}$ distribution is consistent (or, in some cases, \textit{nearly} consistent) with zero to within $3\sigma$ when one or more of the following conditions are met: $f_{\mathrm{hypo}} \lesssim 0.2$, $(P_{\mathrm{hypo}}/P_{\mathrm{out}})/(P_{\mathrm{out}}/P_{\mathrm{2nd-out}}) \gtrsim 3$, $R_{p,\mathrm{hypo}}/R_{p,\mathrm{out}} \lesssim 0.5$, or $\gamma_{\sigma_i'} \gtrsim 6$. These thresholds are both approximate and conservative. The conditions on $P_{\mathrm{hypo}}$ and $R_{p,\mathrm{hypo}}$ indicate that the systems would require a significant deviation from the typical ``peas-in-a-pod'' architectures. Moreover, the condition on $f_{\mathrm{hypo}}$ indicates that only $\sim20\%$ of systems can host an additional outer planet with the expected architecture if the population is to be consistent with the observations. However, we note that the four conditions mentioned above could be relaxed if multiple parameter variations were considered simultaneously (e.g. both larger period ratios and smaller radii). 

The condition on the $\sigma_i'$ scaling factor, $\gamma_{\sigma_i'}$, indicates that our inclination dispersion estimates would have to be very off (by more than a factor of $\sim 6$) before the distribution of $N_{\mathrm{detect}}$ is consistent with zero detections at $3\sigma$. However, this raises questions about how large $\sigma_i'$ can actually be before it becomes too unlikely that the \textit{observed planets} are still co-transiting. In other words, perhaps we can rule out $\sigma_i'$ values that are as large as those required to significantly shift the $N_{\mathrm{detect}}$  distribution. We investigate this using synthetic trials in which we systematically increase $\gamma_{\sigma_i'}$ and determine how frequently the number of transiting planets equals the observed number. For 15 equally-spaced values of $\gamma_{\sigma_i'}$, we consider each Kepler 4+ system, and we reassign each observed planet in the system a random sky-plane inclination according $i\sim\mathcal{N}(\overline{i},\gamma_{\sigma_i'}\sigma_i')$. We then determine how many planets are transiting, $N_{\mathrm{sim}}$, and compare to the observed number of transiting planets, $N_{\mathrm{obs}}$. We calculate the fraction of all 64 systems for which $N_{\mathrm{sim}} = N_{\mathrm{obs}}$. Finally, we repeat these random trials 100 times.  

\begin{figure}
\centering
\epsscale{1.2}
\plotone{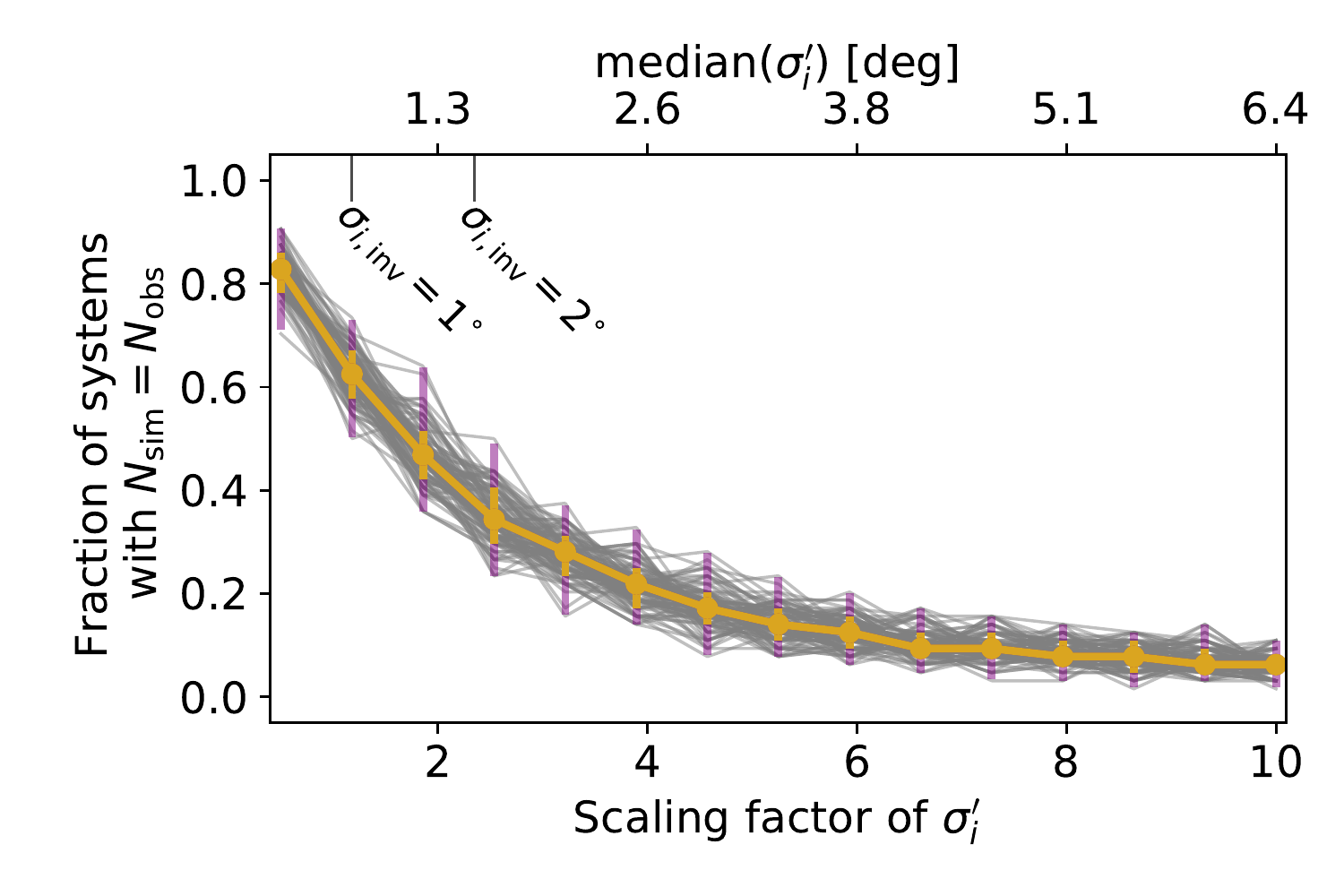}
\caption{Results of simulations that examine the impact of varying $\gamma_{\sigma_i'}$, the scaling factor of $\sigma_i'$, on the transit multiplicity of the observed planets in the Kepler 4+ systems. The y-axis shows the fraction of the 64 Kepler 4+ systems for which the simulated number of transiting planets, $N_{\mathrm{sim}}$, is equal to the observed number of planets in the system, $N_{\mathrm{obs}}$. The gray lines show the individual results from 100 simulations for each value of $\gamma_{\sigma_i'}$. The yellow (purple) lines show the median and the $1\sigma$ ($3\sigma$) intervals. Benchmark values of the dispersion of the inclinations with respect to the invariable plane, $\sigma_{i,\mathrm{inv}}\approx1.6\sigma_i$ (Section \ref{sec: transit chord ratio method validation using SysSim}), are also labeled.}
\label{fig: scaling factor simulations}
\end{figure}

Figure \ref{fig: scaling factor simulations} shows the results of these simulations. The fraction of Kepler 4+ systems for which $N_{\mathrm{sim}} = N_{\mathrm{obs}}$ rapidly decreases as a function of $\gamma_{\sigma_i'}$. Note that the fraction is not unity for $\gamma_{\sigma_i'} = 1$ because the process of resampling the inclinations makes it unlikely to recover all of the exact same planets in transit, given that the observed sample is conditioned upon the planets transiting. With $\gamma_{\sigma_i'}$ as large as $\sim 6$ (the value required earlier), less than $15\%$ of systems would still have as many transiting planets as the observed systems. We can thus safely rule out the possibility that we have underestimated $\sigma_i'$ by a factor as large as required to make the $N_{\mathrm{detect}}$ distribution consistent with zero. However, this still leaves open the possibility that the edge-of-the-multis discrepancy is explained by a substantial increase in the mutual inclinations at long periods \citep[e.g.][]{2021MNRAS.505.1293S}.   

\section{Discussion}
\label{sec: discussion}

\subsection{Constraints on the truncation location}

The hypothetical planet experiments from the previous section indicate that the outer edges of the observed Kepler 4+ systems likely cannot be sculpted by geometric and detection biases alone. Rather, these systems provide evidence either for an average truncation (i.e. occurrence rate drop-off) in the underlying architectures and/or some breakdown of the ``peas-in-a-pod'' patterns at larger orbital separations. We have not yet discussed where this transition occurs, since our hypothetical planet experiments were primarily designed to assess whether it exists rather than where it is. Even so, these experiments can still provide us with some insight.

\begin{figure}
\centering
\epsscale{1.1}
\plotone{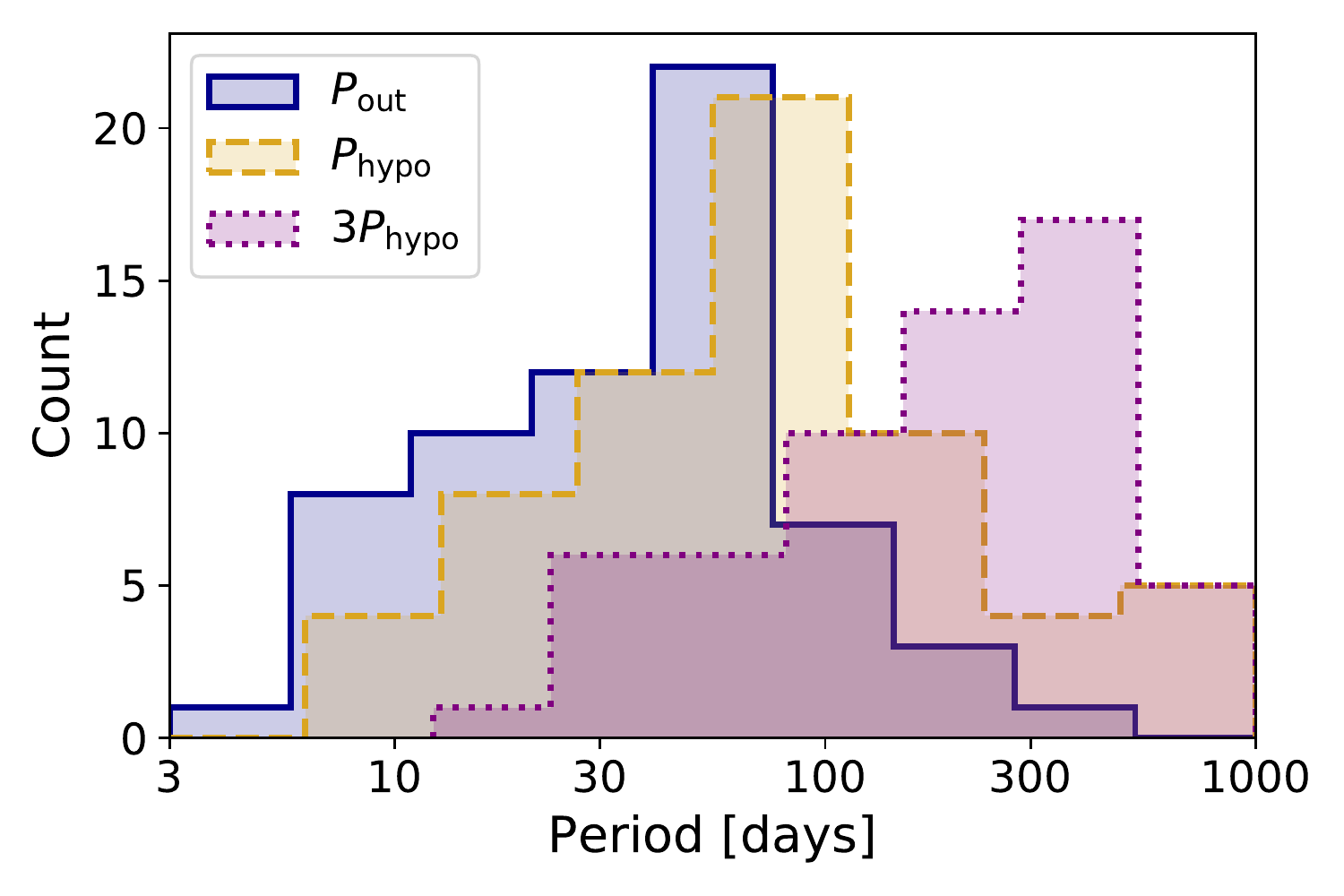}
\caption{Distributions of orbital periods of outermost planets. Blue/solid: orbital periods of the outermost observed planets in the 64 Kepler 4+ systems. Yellow/dashed: orbital periods of the 64 hypothetical planets assuming $P_{\mathrm{hypo}}/P_{\mathrm{out}} = P_{\mathrm{out}}/P_{\mathrm{2nd-out}}$. Purple/dotted: orbital periods of the 64 hypothetical planets assuming $P_{\mathrm{hypo}}/P_{\mathrm{out}} = 3P_{\mathrm{out}}/P_{\mathrm{2nd-out}}$ (which corresponds to the yellow distribution scaled by a factor of 3). These distributions have not been corrected for completeness.} 
\label{fig: Outermost_planet_period_distribution}
\end{figure}

Figure \ref{fig: Outermost_planet_period_distribution} shows the distributions of $P_{\mathrm{out}}$, $P_{\mathrm{hypo}}$ assuming the default ``peas-in-a-pod'' expectation ($P_{\mathrm{hypo}}/P_{\mathrm{out}} = P_{\mathrm{out}}/P_{\mathrm{2nd-out}}$), and $P_{\mathrm{hypo}}$ assuming larger period ratios ($P_{\mathrm{hypo}}/P_{\mathrm{out}} = 3P_{\mathrm{out}}/P_{\mathrm{2nd-out}}$), which corresponds to the approximate condition for which the $N_{\mathrm{detect}}$ distribution would be consistent with observations according to our targeted sampling experiments (Section \ref{sec: Results from targeted sampling}). The medians of the three distributions are 40.6 days, 78.9 days, and 236.6 days, respectively. The distribution of $P_{\mathrm{out}}$ shows a steep drop-off at around $\sim100$ days. Meanwhile, the third distribution rises beyond $\sim100$ days and peaks around $\sim300$ days. Since the third distribution corresponds to a set of hypothetical planets that would be consistent with zero detections to within $3\sigma$, this indicates that the average edge must occur somewhere within the range $P_{\mathrm{edge}} \approx 100-300$ days, or $a_{\mathrm{edge}} \approx 0.5-1$ AU.

\subsection{Theoretical interpretations}
\label{sec: theoretical interpretations}

Whether our results are consistent with an average truncation in the underlying architectures or some breakdown of the ``peas-in-a-pod'' patterns, these findings have important implications for theories of the formation and dynamical evolution of compact multis. The theoretical interpretations will differ depending on which effect is dominant, but the same physical mechanism could, in principle, lead to both an occurrence rate decrease and a change in the architectures at larger separations. Nevertheless, we discuss outer truncation and a ``peas-in-a-pod'' breakdown separately below.

\subsubsection{Outer truncation}
\label{sec: outer truncation}

With regard to outer truncation, one possibility is that, if compact multis form via orbital migration, they may often experience migration traps that prevent inward migration of some planets and cause large gaps in the systems at $\sim100-300$ days (\citealt{2022arXiv220205342Z}; see also e.g. \citealt{2010MNRAS.401.1950P}, \citealt{2014MNRAS.445..479C}, \citealt{2015A&A...575A..28B}, \citealt{2017MNRAS.470.1750I}, \citealt{2019MNRAS.486.3874C}). Recently, \cite{2022arXiv220205342Z} outlined that migration traps should occur in a time-evolving range of planet mass and semi-major axis (e.g. $M_p\sim 0.5-2 \ M_{\oplus}$, $a \lesssim 2$ AU at $\sim5$ Myr) where the co-rotation torque dominates over the Lindblad torque and planets thus migrate outwards instead of inwards. Planets in this mass and semi-major axis range are prevented from reaching the inner regions of the disk, whereas more massive planets avoid the traps and end up on close-in and dynamically-cold orbits. Although \cite{2022arXiv220205342Z} studied this theoretical framework in the context of possible explanations for the ``Kepler dichotomy’’ \citep[e.g.][]{2011ApJS..197....8L, 2012ApJ...758...39J}, it is relevant that their simulations naturally yield planetary systems that are bifurcated into two clusters with a large gap at $\sim100-300$ days, since this is precisely consistent with our observational findings. 

Another physical mechanism that may cause outer architecture truncation is a proposed channel of compact multi formation called “inside-out planet formation” \citep{2014ApJ...780...53C, 2015ApJ...798L..32C}. In this framework, planets are formed sequentially from successive gravitationally unstable rings of $\sim$cm-m sized pebbles that drift inwards via gas drag and build up at the pressure maximum associated with the dead-zone boundary, which separates the inner and outer regions of the disk where the magnetorotational instability is and is not active.\footnote{Other theories \citep[e.g.][]{2021NatAs.tmp..262I} have proposed that planet formation may occur within rings of planetesimals created by other types of pressure maxima in the disk, such as the bump associated with the silicate sublimation line.} Once a planet forms out of the pebble ring, it may migrate or isolate itself from the accretion flow, after which the dead-zone boundary retreats and the process repeats. The process is limited by the finite extent of the dead-zone boundary's retreat and by the mass reservoir that remains available for planet formation after the initial onset \citep{2018ApJ...857...20H}. Accordingly, this model predicts that planet formation will halt at some orbital separation, creating a break in the outer regions of compact multis. 

In contrast to disk effects at the formation epoch, the outer edges of compact multis may also be limited by exterior perturbing planets. Recent results indicate that distant giant planets (with $a \gtrsim 1$ AU, $M_p \gtrsim 0.5 \ M_{\mathrm{Jup}}$) are common in systems with inner super-Earths \citep{2018AJ....156...92Z, 2019AJ....157...52B}. These distant giant planets are often eccentric and may be mutually inclined with respect to the inner planets \citep{2013ApJ...767L..24D, 2020AJ....159...38M}. Several authors have shown that distant giant planets can dynamically excite an inner system of super-Earths, provided that the distant perturber exerts a stronger gravitational influence on the inner planets than their mutual gravitational coupling \citep[e.g.][]{2017AJ....153...42L, 2018MNRAS.478..197P, 2019MNRAS.482.4146D, 2020AJ....160..105S, 2021AJ....162..220T}. The strength of the distant giant planet's perturbation relative to the inner planets' coupling depends on the separation between the outer perturber and the inner system (among other parameters). Accordingly, planets in compact multis that extend out to sufficiently large orbital separations may be dynamically excited or even destabilized by a distant giant. However, since distant giants are present in only a fraction of compact multi systems, their presence likely cannot be the sole explanation of our findings.

%(*Other dynamical instabilities? Maybe like the secular resonances in the asteroid belt?*)

\newpage
\subsubsection{Breakdown of the ``peas-in-a-pod'' patterns}

A different set of theories may be relevant in the case where our results are explained by smaller and/or more widely spaced planets at larger orbital separations. One possibility is that the dominant channel of compact multi formation -- whether it is \textit{in situ} accretion \citep[e.g.][]{2012ApJ...751..158H, 2013ApJ...775...53H, 2013MNRAS.431.3444C, 2015MNRAS.453.1471D}, distant assembly followed by inward migration \citep[e.g.][]{2007ApJ...654.1110T, 2014A&A...569A..56C, 2014MNRAS.445..479C, 2017MNRAS.470.1750I, 2019MNRAS.486.3874C}, or some variation thereof \citep[e.g.][]{2014ApJ...780...53C} -- might naturally produce smaller planet masses at larger orbital separations ($\gtrsim 0.5-1$ AU) due to limitations of solid material or growth timescales.

For instance, in the planet formation simulations by \cite{2017MNRAS.470.1750I}, in which sub-Neptune systems are produced through inward migration and disruption of resonant chains, the resulting systems appear to have a gap at around $\sim0.5-1$ AU, with the planets beyond the gap being less massive ($\lesssim 1 \ M_{\oplus}$). This is also seen in other works such as \cite{2022arXiv220205342Z} and is related to the slower migration timescales for the smaller mass planets. These small planets can also encounter migration traps where the co-rotation torque is dominant (as discussed in Section \ref{sec: outer truncation}), whereas the larger mass planets avoid the traps and migrate inwards. This process naturally leads to a bifurcation of the system, with larger sub-Neptunes inside of $\sim0.5-1$ AU and smaller planets beyond that.

Stepping back from the specific details of planet formation physics, one can consider the general process of energy optimization at the assembly epoch, which also leads to the prediction that the ``peas-in-a-pod'' patterns should break down at larger orbital separations.  \cite{2020MNRAS.493.5520A} (see also \citealt{2019MNRAS.488.1446A}) calculated the lowest energy states available to forming planetary pairs subject to conservation of angular momentum, constant total mass, and fixed orbital spacing. They found that a configuration with approximately equal masses (i.e. ``peas-in-a-pod'') is the most energetically favorable when the total mass in the planets is less than some critical threshold, $m_c \sim 40 \ M_{\oplus}$. Above this threshold, the optimum state is one in which most of the mass is in one planet. The critical threshold $m_c$ decreases with increasing $a$, which indicates that mass uniformity is no longer energetically favorable at sufficiently large orbital distances ($\gtrsim0.5-1$ AU). Thus, the breakdown of the ``peas-in-pod'' patterns might simply be a consequence of planets optimizing the available energy.

To summarize, a variety of physical processes may be responsible for an outer architecture truncation or transition. It is possible that the same physical mechanism (e.g. migration traps) can cause multiple effects (e.g. occurrence rate decrease and smaller planets beyond $\sim0.5-1$ AU). We note that the list of theories above is by no means exhaustive, and it is beyond the scope of this paper to investigate all possibilities in detail. This topic is ripe for future theoretical investigations.

\section{Conclusion}
\label{sec: conclusion}

The inner and outer edges of compact multi-planet systems are fundamental signatures of their formation and evolution. The distribution of the innermost planet orbital periods of the \textit{underlying} distribution of compact multis (i.e. after accounting for observational biases) peaks around $\sim 10$ days \citep{2018AJ....156...24M}, consistent with being a relic from the protoplanetary disk inner edges \citep[e.g.][]{2007ApJ...654.1110T, 2017ApJ...842...40L}. The outer edges are significantly more affected by observational biases and thus have received no prior in-depth investigations, as far as we're aware. However, these biases are well-understood, meaning that a robust characterization is possible. 

In this paper, we presented evidence that the outermost planets in the observed Kepler high-multiplicity systems truncate at smaller orbital periods than expected from geometric and detection biases alone. We showed this using experiments of the detectability of ``hypothetical planets'' orbiting beyond the outermost observed planets in Kepler high-multiplicity systems. These experiments were first demonstrated heuristically in Section \ref{sec: heuristic calculation} and then developed robustly in the remainder of the paper.

%To compute the detectability of hypothetical additional outer planets, we required models for two key components: (1) the geometric transit probability, and (2) the detection efficiency. Our calculations of the transit probabilities used estimates of the mean sky-plane inclination and sky-plane inclination dispersion of each system, which we derived through a novel method involving pairwise transit chord length ratios. Our calculations of detection efficiency were based on the state-of-the-art model from \cite{2019AJ....158..109H}, which incorporates Kepler DR25 data products. We also modified the model using results from \cite{2019MNRAS.483.4479Z} to account for the reduced detection efficiency of planets in high-multiplicity systems. We validated the accuracy of our transit and detection probability calculations using populations of simulated planets from the ``SysSim'' forward model \citep{2018AJ....155..205H, 2019MNRAS.490.4575H, 2020AJ....160..276H}, generated within the framework of the ``maximum AMD model'' \citep{2020AJ....160..276H}.

We considered one hypothetical outer planet per system for 64 Kepler systems with four or more observed planets, with each hypothetical planet's properties dictated by expectations from the ``peas-in-a-pod'' patterns \citep{2018AJ....155...48W, 2017ApJ...849L..33M}. Using models of the transit and detection probabilities, we estimated that $22.11 \pm 3.25$ (approximately $\sim35\%$) of these 64 planets would be transiting and detectable, constituting a $\sim 7\sigma$ difference with zero detections. This is significantly different than analogous results using simulated planetary systems, which were consistent with zero to within $\sim 2-3\sigma$. We thus identified a strong discrepancy between expectations based on hypothetical planets and the lack of additional outer planet detections in the observed systems. Crucially, these results are a lower limit on the true extent of the discrepancy, since we would expect more detections if we assigned more than one hypothetical planet per system.

The discrepancy reveals that there is some truncation or transition in the underlying system architectures at roughly $\sim100-300$ days. There are four distinct possibilities, enumerated below:
\begin{enumerate}
\item There is an average truncation in the underlying architectures. If the ``peas-in-a-pod'' patterns continue to larger separations than observed, consistency with the observations would require that fewer than $\sim20\%$ of systems host additional planets at the next outermost expected periods.
\item The period ratios increase at larger orbital separations. This would require an average increase in the period ratios (relative to the ``peas-in-a-pod'' expectations) by a factor of $\gtrsim3$. 
\item The planet radii are smaller at larger orbital separations. This would require an average decrease in the radii (relative to the ``peas-in-a-pod'' expectations) by a factor of $\lesssim0.5$.
\item The mutual inclinations increase at larger orbital separations. This would require an average increase in the inclination dispersions by a factor of $\gtrsim 6$.
\end{enumerate}

Whether the outer architectures of the compact multis have some truncation or a breakdown of the ``peas-in-a-pod'' patterns, these results have important consequences for theories of planet formation and dynamical evolution. Several physical processes could potentially give rise to these features (Section \ref{sec: theoretical interpretations}). However, one particularly complementary prediction from the migration model of close-in sub-Neptune formation is that planets in a certain mass and semi-major axis range should experience migration traps that prevent inward migration \citep[e.g.][]{2017MNRAS.470.1750I}, leading to a gap in the resulting systems at $\sim 100-300$ days and the existence of smaller mass planets outside of this gap \citep{2022arXiv220205342Z}. We encourage future work on this and other theoretical processes pertinent to the outer regions of compact multis. 

With the Kepler dataset alone, it is impossible to determine whether our results provide evidence for a truncation or some other transition at $\gtrsim 0.5-1$ AU. However, future observations may characterize the outer architectures enough to distinguish these possibilities. In particular, extreme precision radial velocity observations \citep[e.g.][]{2021A&A...645A..96P} may allow for a detailed understanding of the outer architectures of some compact multis orbiting particularly bright or quiet stars, with systems like TOI-178 \citep{2021A&A...649A..26L} being a prime example. Long-term monitoring of transit timing variations can also signal the presence of more distant, non-transiting planets \citep[e.g.][]{2012Sci...336.1133N}. Future high-precision photometric monitoring, especially with the PLATO mission  \citep{2014ExA....38..249R}, will shed more light on the outer architectures of compact multis by revisiting the Kepler field and potentially targeting some new fields for multi-year baselines. Finally, continued efforts at characterization of gas giants at larger ($\sim$few-AU) separations \citep[e.g.][]{2021ApJS..255....8R, 2021ApJS..255...14F}, especially in systems with inner sub-Neptunes \citep[e.g.][]{2022ApJ...926...62C}, will gradually provide a more holistic understanding of inner and outer system interactions and dynamics.

\section{Acknowledgements}
We thank the anonymous referee for their careful review and constructive comments. We are grateful to Fred Adams, Chris Burke, Eric Ford, Dan Tamayo, Jonathan Tan, and Josh Winn for helpful discussions. S.C.M. was supported by NASA through the NASA Hubble Fellowship grant \#HST-HF2-51465 awarded by the Space Telescope Science Institute, which is operated by the Association of Universities for Research in Astronomy, Inc., for NASA, under contract NAS5-26555. This research has made use of the NASA Exoplanet Archive, which is operated by the California Institute of Technology, under contract with the National Aeronautics and Space Administration under the Exoplanet Exploration Program.

\bibliographystyle{aasjournal}
\bibliography{main}

\end{document}